\documentclass[]{jfm}

\usepackage{graphicx}
\usepackage{newtxtext}
\usepackage{newtxmath}
\usepackage{natbib}
\usepackage{mathtools}
\usepackage{hyperref}
\hypersetup{
    colorlinks = true,
    urlcolor   = blue,
    citecolor  = black,
}

\newcommand{\RomanNumeralCaps}[1]
\linenumbers
\usepackage{lscape}
\usepackage{caption}
\usepackage{xcolor}
\usepackage{placeins}

\definecolor{darkred}{HTML}{8B0000}   

\title{Turbulence and added drag over acoustic liners}

\author{Haris Shahzad
  \corresp{\email{h.shahzad@tudelft.nl}},
  Stefan Hickel
 \and Davide Modesti}

\affiliation{Aerodynamics Group, Faculty of Aerospace Engineering, Delft University of Technology,Kluyverweg 2, 2629 HS Delft, The Netherlands}

\begin{document}

\maketitle

\begin{abstract}
We present pore-resolved direction numerical simulations (DNS) of turbulent flows
grazing over perforated plates, that closely resemble the acoustic liners found on aircraft engines.
Our DNS explore a large parameter space including the effects of porosity, thickness, and viscous-scaled diameter of the perforated plates,
at friction Reynolds numbers $Re_\tau = 500$--$2000$,
which allows us to develop a robust theory for estimating the added drag induced by acoustic liners. 
We find that acoustic liners can be regarded as porous surfaces with a wall-normal permeability and that
the relevant length scale characterizing their added drag is the inverse of the wall-normal Forchheimer coefficient.
Unlike other types of porous surfaces featuring Darcian velocities inside the pores, the flow inside the orifices of
acoustic liners is fully turbulent, with a magnitude of the wall-normal velocity fluctuations comparable to the peak in the near wall cycle.
We provide clear evidence a fully rough regime for acoustic liners, also confirmed by the increasing relevance of pressure drag.
Once the fully rough asymptote is reached, 
canonical acoustic liners provide an added drag comparable to sand-grain roughness with viscous-scaled height 
matching the inverse of the viscous-scaled Forchheimer permeability of the plate.
\end{abstract}


\section{Introduction}
\label{sec:introduction}

\begin{figure}
    \begin{center}
        \includegraphics[scale=0.18]{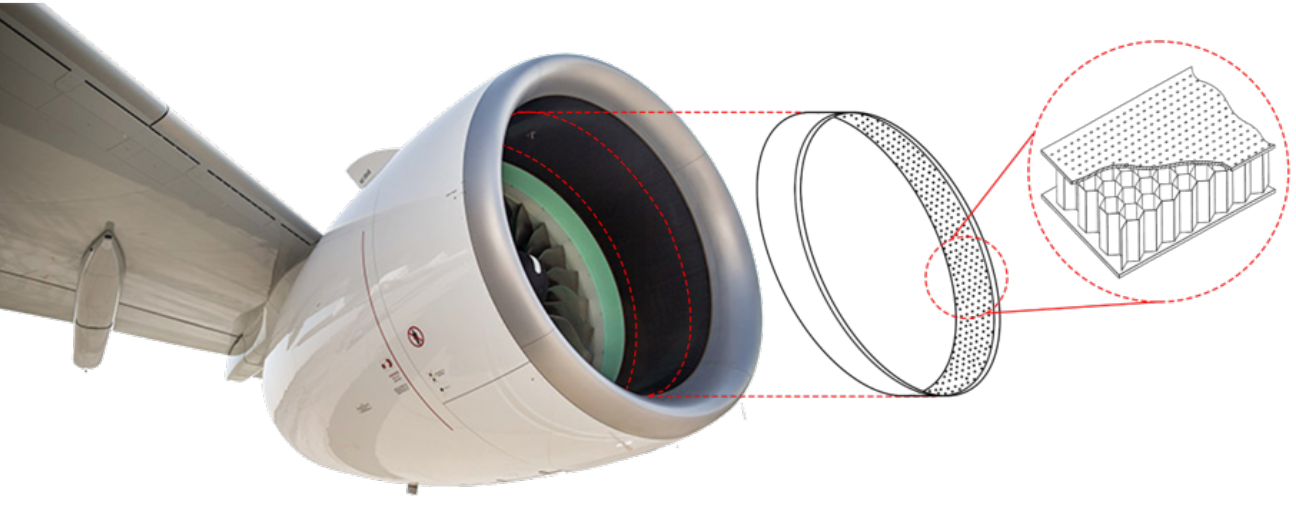}
        \put(-240,110){(\textit{a})}
        \includegraphics[scale=0.03]{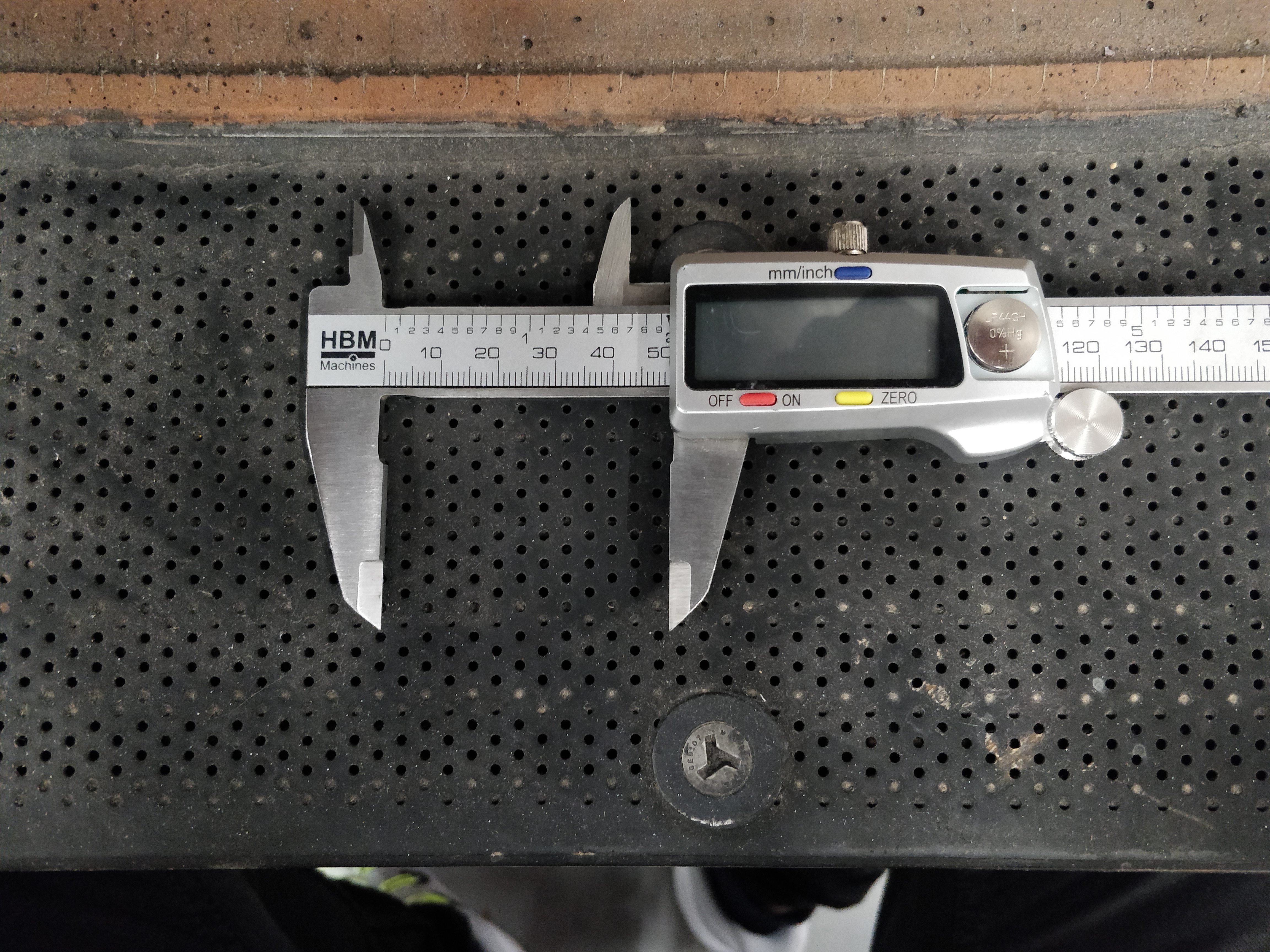}
        \put(-140,110){(\textit{b})}
        \caption{
         (\textit{a}) Turbofan engine of a civil aircraft with acoustic liners
            on the air intake.
            (\textit{b}) The typical pore size of acoustic liners used in turbofan engines.}
            \label{fig:liner}
    \end{center}
\end{figure}

\begin{figure}
    \centering
        \includegraphics[width = 1\textwidth] {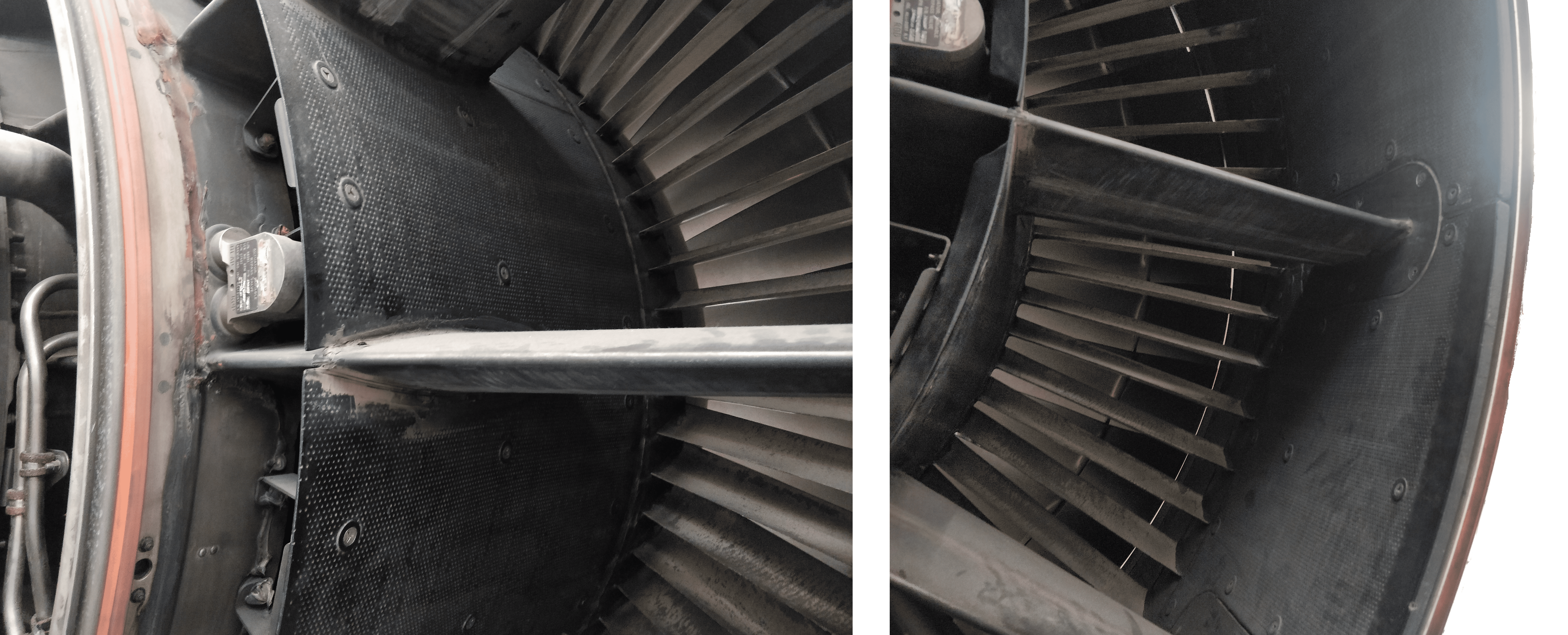} 
    \caption{Acoustic liners around the fan of a turbofan engine.}
    \label{fig:picliner}
\end{figure}

Aircraft engines are the primary source of noise during take-off and landing. 
In order to meet noise regulations, the nacelle of modern engines is coated with acoustic liners, 
which represent the state-of-the-art technology for engine noise abatement. 
Acoustic liners are panels with a sandwich structure, consisting of a honeycomb core, bounded by a perforated facesheet and a solid backplate, see figure~\ref{fig:liner}(\textit{a}).
They cover the nacelle inner surface, both in front of the fan and in the by-pass duct (see figure \ref{fig:picliner}),
and can theoretically absorb all incoming sound if the resonant frequency of the liner is tuned to the frequency of the incoming acoustic 
wave~\citep{hughes_absorption_1990, dowling_sound_1992,kirby_impedance_1998}. 
In realistic conditions, several studies have shown that acoustic liners reduce fan noise as much as 3--6 $dB$ \citep{casalino_turbofan_2018,shur_unsteady_2020}.
They are, therefore, an essential part of aircraft engines.

Although the sound attenuation mechanism is well understood, the aerodynamic characteristics of these surfaces are less clear.
Several authors agree that liners increase aerodynamic drag as compared to a hydraulically smooth wall~\citep{wilkinson_influence_1983,howerton_acoustic_2015,jasinski_mechanism_2020}.
However, an extensive literature study summarized in table~\ref{tab:dragstudies}, shows that 
reported values for the actual drag increase caused by acoustic liners vary between 2\% and 500\%. 
Hence, at present, we lack a theory for the prediction of the aerodynamic drag over acoustic liners.

\citet{wilkinson_influence_1983} was amongst the first to perform experiments of turbulent boundary layers 
over porous plates for different values of the viscous-scaled orifice diameter $d^+\coloneqq d/\delta_v$,
viscous-scaled plate thickness $t^+\coloneqq t/\delta_v$ and plate porosity (open-area ratio) $\sigma$.
Here, $\delta_v=\nu_w/u_\tau$ is the viscous length scale, $\nu$ the kinematic viscosity of the fluid, $u_\tau=\sqrt{\tau_w/\rho_w}$ the friction velocity,
$\tau_w$ the drag per plane area, $\rho$ the fluid density and the subscript $w$ denotes quantities evaluated at the wall.

More recently, several experiments have been
conducted in the Grazing Flow Impedance Tube (GFIT) facility at NASA~\citep{jones_design_2004}
and considerable effort has been dedicated to estimating the added drag provided by acoustic liners
using a static pressure drop approach~\citep{howerton_acoustic_2015,howerton_acoustic_2016,howerton_conventional_2017}.
These experimental campaigns considered several liner geometries, for both conventional and more exotic
configurations~\citep{howerton_acoustic_2015,howerton_conventional_2017}, and reported a drag increase between 16\% and 350\% compared to a smooth wall.

An important finding of the GFIT experiments is that the cavity depth has a negligible contribution
to the total drag, which instead is largely influenced by the orifice diameter, plate porosity and the facesheet thickness.
For instance, \citet{howerton_acoustic_2015} noted that, for constant porosity, reducing the diameter of the orifices reduced drag. Similarly, \citet{howerton_conventional_2017} reported 50\% drag increase for porosity
$\sigma=0.08$ and 400\% drag increase for $\sigma=0.3$, for the same flow conditions and approximately the same orifice diameter. Additionally, these experiments suggest that it is possible to reduce the drag penalty without
harming the noise attenuation.

\citet{gustavsson_correction_2019} performed experiments over several acoustic liner geometries
and reported a drag increase between $30\%$--$50\%$, without incoming acoustic waves, arguing
that the added drag might be even larger in the presence of incoming noise.

Numerical simulations of turbulent flows over acoustic liners are also available, 
but very often they rely on simplified configurations or wall models because pore-resolved simulations
are computationally expensive.
A common approach that has been pursued for reducing the computational cost is to simulate
a single cavity rather than an array of resonators~\citep{zhang_numerical_2011,zhang_numerical_2016,avallone_lattice-boltzmann_2019}.
\citet{zhang_numerical_2016} performed direct numerical simulation (DNS)
of turbulent grazing flow over a single resonator with a similar cavity geometry 
as the one studied by \citet{howerton_acoustic_2015} in the GFIT ~\citep{jones_comparison_2004}.
However, the simulations were at a much lower friction Reynolds number, see table~\ref{tab:dragstudies}.
For a free stream Mach number $M_\infty=0.5$, \citet{zhang_numerical_2016} reported a minor drag increase of $4.2\%$ with respect to a smooth wall
in the absence of sound waves, whereas they found a drag increase of about 25\% when including sound waves with 
an intensity of $140dB$.
These results seem to contradict experiments of \citet{howerton_acoustic_2015}  who reported a drag increase of about 50\% both with and without incoming sound waves, at matched Mach number
and cavity geometry.
This discrepancy can probably be traced back to the simplified numerical setup wherein only one single orifice is
simulated, resulting in a very low porosity $\sigma=0.0099$, compared to $\sigma=0.08$ in the experiments.

Another common simplification in numerical simulation is to approximate the effect of acoustic liners with an equivalent 
impedance boundary condition~\citep{tam_time-domain_1996}, which substantially reduces
the computational cost. 
However, the accuracy with which the impedance boundary condition represents the real acoustic liner geometry
is not well understood and discrepancies can be observed in literature. 
For instance, \citet{olivetti_direct_2015} performed DNS of turbulent channel flow with impedance boundary conditions and
did not report changes in the structure of the near wall turbulence. 
On the contrary, \citet{scalo_compressible_2015} and \citet{sebastian_numerical_2019} performed large eddy simulations of turbulent channel flow with a characteristic impedance boundary condition ~\citep{fung_impedance_2000, fung_time-domain_2004}, 
and noted significant changes in the structure of the near wall cycle which could, for some cases, be completely replaced by
Kelvin--Helmholtz-like rollers, with drag increase up to 500\%. 

Despite the very large discrepancies between previous studies,
there seem to be a consensus that the added drag depends both on the orifice diameter
$d$ and the porosity of the facesheet $\sigma$.
This type of functional dependency has been observed
in turbulent grazing flows over porous substrates, which
is a hint that acoustic liners might be regarded
as porous surfaces, permeable in the wall-normal direction.
Porous surfaces differ from other types of impermeable surface textures
such as roughness.
Flows over rough surfaces are characterized by the pressure drag induced by the topography~\citep{leonardi_direct_2003,leonardi_channel_2010,chung_predicting_2021}.
When pressure drag dominates over viscous drag, the skin-friction coefficient becomes independent of the Reynolds number, a regime that
we denote as `fully rough'~\citep{leonardi_direct_2003,chung_predicting_2021}.
The flow through many common porous surfaces (such as metal foams, sedimentary rocks, sandstone, conglomerates)
can be characterized by the permeability of the medium $K_i$, namely the
ease with which the flow in $i$ direction passes through the porous surface.
Our current understanding of porous surfaces is that pressure drag and permeability are intrinsically coupled
because the grazing flow perceives the pores as a non-smooth surface texture (which provides pressure drag), but it is also able to penetrate 
into the pores. 
Therefore the flow has a `roughness' component (pressure drag) and a `porous' component (permeability), that are inseparable. 

Some authors attempted to separate these two concurring effects.
For instance, \citet{manes_turbulence_2009} studied the similarities and differences between roughness and porous surfaces
by carrying out experiments of (impermeable) rough walls with the same surface topography of porous substrates, in the attempt
to isolate the effect of the permeability.
They found that permeability largely contributes to the total drag and a fully rough regime did not emerge for the permeable substrate.
\citet{esteban_2022} carried out experiments of permeable surfaces and delineated the effects of roughness and permeability by considering permeable surfaces with constant permeability but different thicknesses.
They found that changing the thickness altered the added drag and such an affect could be attributed to the 'roughness' component of the geometry.
Unlike \citet{manes_turbulence_2009}, \citet{esteban_2022} found that porous surfaces do indeed approach a fully rough regime.
\citet{breugem_influence_2006} carried out DNS of porous surfaces by modelling the substrate with a Darcy boundary condition.
The authors pointed out that the duality between `rough' and `porous' surface
is reflected in the presence of three concurring length scales,
namely the boundary layer thickness $\delta$, the pore diameter $d$, and the square root of the permeability $\sqrt{K_i}$.
These length scales can be converted into the friction Reynolds number $\Rey_\tau=\delta/\delta_v$, 
the viscous-scaled pore diameter $d^+=d/\delta_v$ and the viscous-scaled square root of the permeability $\sqrt{K_i}^+=\sqrt{K_i}/\delta_v$.
The authors reason that the effect of the ``roughness'' and ``porous'' components are separated if there is enough separation between
the two length scales while having $d<<\delta$.
These conditions are somehow always assumed by models such as Darcy's boundary conditions~\citep{breugem_influence_2006,rosti_direct_2015,rosti_turbulent_2018}
and by impedance boundary conditions~\citep{scalo_compressible_2015}. This essentially corresponds to a surface with small pores $d^+<<70$, but very high porosity $\sigma$ (i.e. open area ratio) and therefore high permeability, $\sqrt{K_i}^+>1$.

Acoustic liners do not satisfy these conditions.
Figure~\ref{fig:liner}(\textit{b}) shows that
the plate porosity is relatively small, typically $\sigma=0.08$--$0.3$ and the orifice diameter is about $2$--$3$mm.
The orifice diameter with respect to the boundary layer length scales can be estimated assuming an aircraft at cruise condition with Mach Number $M_\infty\approx0.8$, velocity $u_{\infty} \approx 240 m/s$,
and kinematic viscosity $\nu \approx 3.5 \times 10^{-5} m^2/s$.
In these conditions, we can estimate a Reynolds number based on the length of the
air intake of about $\Rey_L \approx 10^{7}$, corresponding to friction Reynolds number $\Rey_\tau\approx 6600$,
with boundary layer thickness $\delta\approx 28mm$, and viscous length scale $\delta_v=4\mu m$.
Therefore, acoustic liners in operating conditions have $d/\delta\approx0.07$, $d^+\approx500$.
The depth of a cavity is typically $h = 40mm$, corresponding to $h/\delta \approx 1.5$. 

Hence, acoustic liners have low porosity, but relatively large orifices,
which is the opposite of canonical porous surfaces, which can reach $\sigma>0.8$~\citep{breugem_influence_2006,rosti_direct_2015}.
Therefore, the operating regime of acoustic liners would exclude them from the canonical definition of porous surfaces,
although the drag dependence on the porosity would suggest the opposite. 

This literature survey shows that there have been several attempts to measure the added drag caused by acoustic liners,
both experimentally and numerically, suggesting a large interest of the community in this topic.
However, the discrepancies between previous studies are too large to be acceptable. This large uncertainty
can be associated with the critical modelling assumptions that have been used in numerical simulations
and the difficulty in measuring drag in experiments.
From a more fundamental perspective, it is not clear if acoustic liners can be regarded as porous surfaces
or as surface roughness, because their geometry does not fall in either of these canonical classifications.

In this work, we aim at developing a rigorous theoretical framework to characterize acoustic liners within
the larger body of non-smooth surface textures. We believe that this can only be achieved by performing
pore-resolved DNS, which allows us to have access to the 3D flow field and to accurately measure the drag without
relying on additional modelling assumptions.

\begin{landscape}
\begin{table*}
\centering
\resizebox{0.95\linewidth}{!}{
\begin{tabular}{lccccccccccc}
& & & & & & & & & & &   \\
& & & & & & & & & & &   \\
& & & & & & & & & & &   \\
& & & & & & & & & & &   \\
& & & & & & & & & & &   \\
            & $M_{\infty}$       & $Re_{\delta}$                & $Re_{\tau}$           & $h^+$                 & $h/\delta$     & $d^+$                   & $d/\delta$            & $t^+$               & $t/\delta$       & $\sigma$      & $\Delta D$    \\ 
\hline
\citet{howerton_acoustic_2015}  & $0.1-0.5$  & $ 2.44 \times 10^{5^\circledast}$ & $ 7800^{\circledast}$ & $14200^{\circledast}$ & $1.8^{\circledast}$ & $180-370^{\circledast}$ & $0.025-0.05^{\circledast}$ & $370^{\circledast}$ & $0.05^{\circledast}$  & $0.08$        & $30-50$\%     \\
\hline
\citet{howerton_acoustic_2016}  & $0.3-0.5$  & $ 2.44 \times 10^{5^\circledast}$ & $ 7800^{\circledast}$ & $14200^{\circledast}$ & $1.8^{\circledast}$ & $280^{\circledast}$     & $0.036^{\circledast}$      & $280^{\circledast}$ & $0.036^{\circledast}$ & $0.08$        & $16-20$\%     \\
\hline
\citet{howerton_conventional_2017}  & $0.3-0.5$  & $ 2.44 \times 10^{5^\circledast}$ & $ 7800^{\circledast}$ & $14200^{\circledast}$ & $1.8^{\circledast}$ & $280^{\circledast}$     & $0.036^{\circledast}$      & $280^{\circledast}$ & $0.036^{\circledast}$ & $0.1$         & $10-15$\%     \\
            & $0.3-0.5$  & $ 2.44 \times 10^{5^\circledast}$ & $ 7800^{\circledast}$ & $14200^{\circledast}$ & $1.8^{\circledast}$ & $660^{\circledast}$     & $0.084^{\circledast}$      & $660^{\circledast}$ & $0.084^{\circledast}$ & $0.2$         & $80-130$\%    \\
            & $0.3-0.5$  & $ 2.44 \times 10^{5^\circledast}$ & $ 7800^{\circledast}$ & $14200^{\circledast}$ & $1.8^{\circledast}$ & $470^{\circledast}$     & $0.06^{\circledast}$       & $470^{\circledast}$ & $0.06^{\circledast}$  & $0.3$         & $200-350$\%   \\
\hline
\citet{wilkinson_influence_1983}   & $0$        & $(2.8-6.4) \times 10^{4^\dag}$    & $500-2000^{\dag}$     & $350-1100$            &                     & $40-150$                &                            & $30-150$            &                       & $0.06-0.12$   & $2-20$\%      \\
            & $ 0$       & $(2.8-6.4) \times 10^{4^\dag}$    & $ 500-2000^{\dag}$    & $600-3700$            &                     & $9-55$                  &                            & $6-35$              &                       & $0.047-0.139$ & $30-60$\%     \\
\hline
\citet{gustavsson_correction_2019} & $0.3-0.6$  & $(4.8-8.3) \times 10^{4^\dag}$         & $ 2000-3000^{\dag}$   & $12000-20000^{\dag}$ & $5.85-6.05^{\dag}$   & $350-550^{\dag}$     & $0.15-0.17^{\dag}$                       & $360-600^{\dag}$      & $0.15-0.17^{\dag}$                  & $0.0853$      & $30-50$\%     \\
\hline
\citet{zhang_numerical_2016}       & $0.$       & $ 2.26 \times 10^{4^\dag}$        & -                     & -                     & -                   & $114$                   & $0.05$                     & -                   & -                     & $0.0099$      & $4-100$\%    \\
\hline
\citet{scalo_compressible_2015}       & $0.05-0.5$ & $6900$                            & -                     & \multicolumn{7}{c}{Impedance Boundary Condition}                                                                                                                 & $\leq 325$\%  \\
\hline
\citet{sebastian_numerical_2019}   & $0.3$      & $6900$                            & $400-1000$            & \multicolumn{7}{c}{Impedance Boundary Condition}                                                                                                                 & $\leq 575$\%  \\
\hline
\citet{jimenez_turbulent_2001}     & $0$        & $2830$                            & $180-215$             & \multicolumn{7}{c}{Resistance Boundary Condition ($X=0$)}                                                                                                        & $21-44$\%     \\
\hline
\end{tabular}
}
\captionsetup{width=.95\linewidth}
	\caption{Dataset of previous studies on drag over acoustic liner geometries. $M_{\infty}$ is the Mach number. $Re_{\delta}=u_0\delta/\nu$ is the Reynolds number based on the boundary layer thickness and external velocity (free-stream velocity for boundary layers or bulk flow velocity for channel flow simulations\citep{scalo_compressible_2015,sebastian_numerical_2019,jimenez_turbulent_2001}) and $\Rey_\tau$ is the friction Reynolds number. The liner geometry is defined by the orifice diameter $d$, the depth of the cavity $h$, the thickness of the facesheet $t$ and and the porosity $\sigma$. $\Delta D$ is the percentage increase in drag observed in these studies. Quantities that are approximated are denoted using the $^{\dag} superscript$. Quantities that are approximated with theliner  aid of the RANS simulations of the GFIT by \citet{zhang_numerical_2016} are denoted using the $^{\circledast}$ superscript.}
\label{tab:dragstudies}
\end{table*}
\end{landscape}

\section{Methodology}

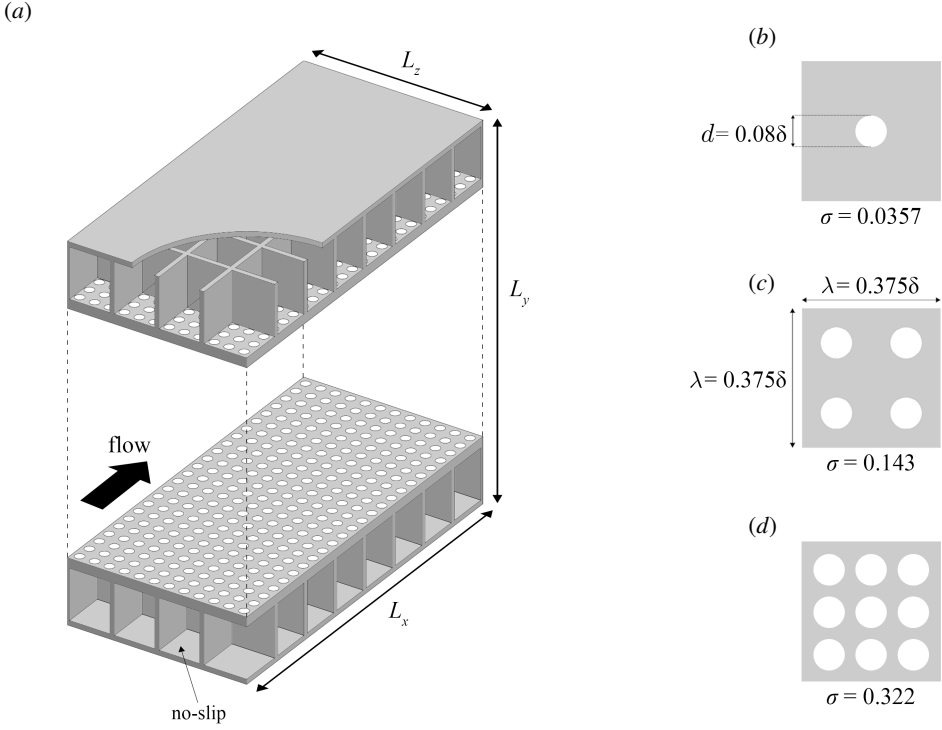
\begin{figure}
    \centering
        \includegraphics[scale = 1] {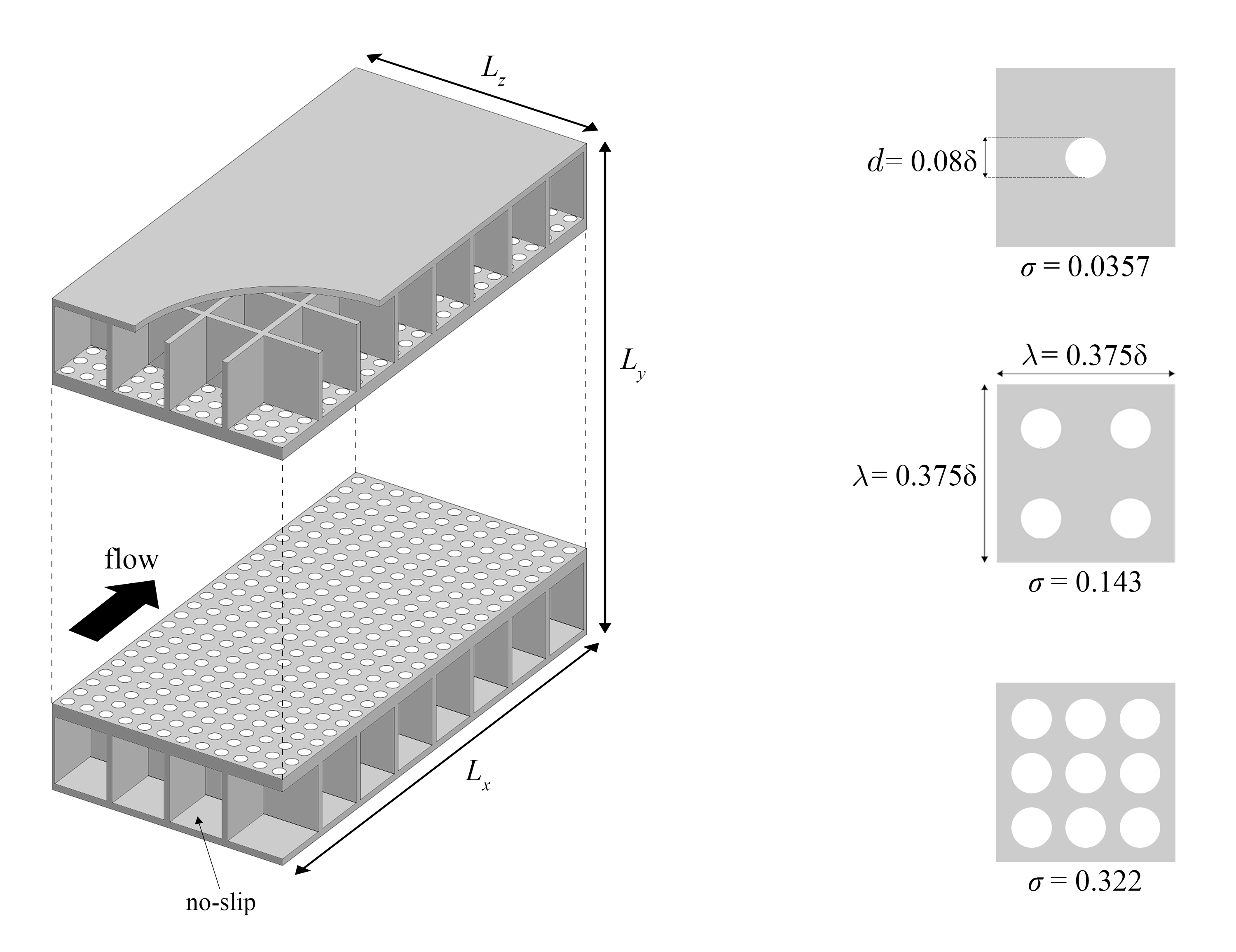} 
    \put(-370,275){(\textit{a})}
    \put( -90,265){(\textit{b})}
    \put( -90,173){(\textit{c})}
    \put( -90, 82){(\textit{d})}
    \caption{Sketch of the computational domain. Turbulent channel flow configuration with box dimension $L_x \times L_y \times L_z $. Different porosities are considered by increasing the number of holes per cavity. The three different porosities, $\sigma$ are shown on the right.}
    \label{fig:instant}
\end{figure}

\label{sec:methodology}

\begin{table}
\centering
\resizebox{1\linewidth}{!}{
\begin{tabular}{cccccccccccccc}
         & $Re_b$ & $Re_\tau$ & $d^+$ & $\sigma$ & $1/\alpha^+$ & $\Delta U^+$ & $C_f \times 10^3$ & $\Delta x^+$ & $\Delta y^+_\text{min}$ & $\Delta z^+$ & $N_x$ & $N_y$ & $N_z$  \\ 
\hline
$S_1$    & 9268   & 506.1     & 0     & 0        & 0            & -            & 4.578             & 5.1          & 0.80                    & 5.1          & 300   & 350   & 150    \\
$S_2$    & 21180  & 1048      & 0     & 0        & 0            & -            & 3.791             & 5.2          & 0.80                    & 5.2          & 600   & 600   & 300    \\
$S_3$    & 45240  & 2060      & 0     & 0        & 0            & -            & 3.201             & 5.2          & 0.80                    & 5.2          & 1200  & 800   & 600    \\
$L_1$    & 9139   & 503.5     & 40.3  & 0.0357   & 0.0528       & 0.14         & 4.598             & 1.1          & 0.80                    & 1.1          & 1500  & 500   & 750    \\
$L_{t1}$ & 9139   & 505.3     & 40.4  & 0.0357   & 0.0287       & 0.17         & 4.738             & 1.5          & 0.81                     & 1.5          & 1000  & 500   & 500    \\
$L_2$    & 8794   & 496.4     & 39.7  & 0.142    & 0.859        & 0.56         & 4.855             & 1.0          & 0.80                    & 1.0          & 1500  & 500   & 750    \\
$L_{t2}$ & 8794   & 515.5     & 41.2  & 0.142    & 0.552        & 0.69         & 4.856             & 1.6          & 0.82                    & 1.6          & 1000  & 500   & 500    \\
$L_3$    & 8264   & 505.3     & 40.4  & 0.322    & 5.14         & 1.90         & 5.539             & 1.0          & 0.81                    & 1.0          & 1500  & 500   & 750    \\
$L_4$    & 19505  & 1038      & 83.0  & 0.142    & 1.727        & 0.96         & 4.363             & 2.1          & 0.83                    & 2.1          & 1500  & 800   & 750    \\
$L_{t4}$ & 19505  & 1047      & 83.8  & 0.142    & 1.120        & 1.40         & 4.475             & 3.1          & 0.84                    & 3.1          & 1000  & 800   & 500    \\
$L_5$    & 17810  & 1026      & 82.1  & 0.322    & 10.4         & 2.78         & 5.058             & 2.1          & 0.82                    & 2.1          & 1500  & 800   & 750    \\
$L_{t5}$ & 17810  & 1055      & 84.4  & 0.322    & 6.692        & 3.28         & 5.317             & 3.1          & 0.84                    & 3.2          & 1000  & 800   & 500    \\
$L_6$    & 35470  & 2044      & 164.0 & 0.322    & 20.8         & 4.44         & 5.267             & 4.1          & 0.82                    & 4.1          & 1500  & 1400  & 750    \\
\hline
\end{tabular}
}
\caption{DNS dataset comprising smooth, $(S_n)$ and liner $(L_n)$ and $(L_{tn})$ cases. Cases $(L_{tn})$ have half the thickness of cases $(L_{n})$. $\sigma$ is the porosity and $\alpha$ is the Forchheimer permeability. $\Delta U^+$ is the Hama roughness function measured at $y^+=100$. $C_f=2/u_\delta^{+2}$ is the skin-friction coefficient, where $u_\delta^+$ is the viscous-scaled velocity at the channel centerline.  Simulations are performed in computational box with dimensions $L_x \times L_y \times L_z = 3\delta \times 2\delta \times 1.5\delta$. $\Delta x^+$ and $\Delta z^+$ are the viscous-scaled mesh spacing in the streamwise and spanwise direction. and $\Delta y^+_{\text{min}}$ is the minimum mesh spacing in the wall normal direction. $d^+$ is the orifice diameter.}
\label{tab:cases}
\end{table}

We solve the compressible Navier--Stokes equations for a calorically perfect gas,

\begin{equation}
    \dfrac{\partial \rho }{\partial t}
  + \dfrac{\partial \rho u_i}{\partial x_i} = 0,
\end{equation}
\begin{equation}
    \dfrac{\partial \rho u_i }{\partial t}
  + \dfrac{\partial \rho u_i u_j}{\partial x_i} = 
  - \dfrac{\partial p}{\partial x_i} 
  + \dfrac{\partial \sigma_{ij}}{\partial x_j}
  + \Pi \delta_{i1},
\end{equation}
\begin{equation}
    \dfrac{\partial \rho E }{\partial t}
  + \dfrac{\partial \rho u_i H}{\partial x_i} = 
  - \dfrac{\partial q_i}{\partial x_i} 
  + \dfrac{\partial \sigma_{ij} u_i}{\partial x_j}
  + \Pi u + \Pi_T,
\end{equation}

where $u_i=\left\{u,v,w\right\}$ are the velocity components, $\rho$ is the density, $p$ is the pressure, $E = c_vT +u_iu_i/2$ is the total energy per unit mass and $H = E + p/\rho$ is the total enthalpy. $c_p$ and $c_v$ are the heat capacities at constant pressure and constant volume. $q_j$ and $\sigma_{ij}$ are the heat flux vector and viscous stress tensor,

\begin{equation}
    \sigma_{ij} = \mu \left(    
      \dfrac{\partial u_i}{\partial x_j}
    + \dfrac{\partial u_j}{\partial x_i}
    - \dfrac{2}{3}\dfrac{\partial u_k}{\partial x_k}\delta_{ij}
    \right),
\end{equation}
\begin{equation}
    q_j = -k \dfrac{\partial T}{\partial x_j},
\end{equation}

where $k = c_p\mu/Pr$ is the thermal conductivity. The Prandtl number is $Pr = 0.72$. 
The viscosity dependence on the temperature is 
accounted for using a power law with exponent $0.75$.  
We consider the plane channel flow configuration wherein
the fully developed flow between two plates is driven in the streamwise direction
by a uniform body force, $\Pi$, which is adjusted every time step to maintain a
constant mass flow rate and the power spent is added to the total energy equation.
A uniform bulk cooling term, $\Pi_T$, is also added to the total energy equation
to maintain a constant bulk flow temperature.
The bulk velocity, temperature and density are defined as,

\begin{equation}
u_b = \dfrac{1}{\rho_b V} \int_{V} \rho u  \; dV, \quad T_b = \dfrac{1}{\rho_b u_b V}  \int_{V} \rho u T \; dV,\quad \rho_b = \dfrac{1}{V} \int_{V} \rho \; dV, 
\end{equation}

where $V=L_x\times2\delta \times L_z$ is the fluid volume between the top 
and bottom perforated plates, see figure~\ref{fig:instant}.

The Navier--Stokes equations are solved using the solver STREAmS~\citep{bernardini_streams_2021}.
The non-linear terms in the Navier--Stokes equations are discretized using an energy-conservative scheme in locally conservative form~\citep{pirozzoli_10}.
The viscous terms are expanded into a Laplacian form and approximated with sixth-order central finite-difference formulas to avoid odd-even decoupling phenomena. Time stepping is carried out using Wray's three-stage third-order Runge--Kutta scheme \citep{spalart_spectral_1991}.

The complexity of the roughness geometry is handled using a ghost-point-forcing immersed boundary method to treat arbitrarily complex geometries~\citep{piquet_comparative_2016,vanna_sharp-interface_2020}.
The geometry of the solid body is provided in OFF format for 3D objects, and the computational geometry library CGAL~\citep{the_cgal_project_cgal_2022} is used to perform the ray-tracing algorithm.
This allows us to define the grid nodes belonging to the fluid and the solid, and to compute the distance of each point from the interface.
To retain the same computational stencil close to the boundaries, the first three layers of interface points inside the body are tagged as ghost nodes.
For each ghost node, we identify a reflected point along the wall-normal, laying inside the fluid domain.
We interpolate the solution at the reflected point using a trilinear interpolation and use the values at the reflected points to fill the ghost nodes inside the body to apply the desired boundary condition.
An extensive description of the algorithm is available in the work by~\citet{vanna_sharp-interface_2020}, and validation of the present implementation is available in the Appendix.

The simulations are carried out in a rectangular box of size $L_x \times L_y \times L_z = 3\delta \times (2\delta+2h) \times 1.5\delta$, where $\delta$ is the channel half-width. 
We use uniform mesh spacing in the streamwise and spanwise directions. 
In the wall-normal direction, the mesh is clustered towards the facesheet walls and coarsened towards the backplate and the channel centre. 
The simulations are performed at bulk Mach number, $M_b = u_b/c_w = 0.3$, where $c_w$ is the speed of sound at the wall. 
The bulk-to-wall temperature ratio is fixed $T_b/T_w = 1$, 
which corresponds to an isothermal cold wall with $T_w/T_{aw} = 0.984$, where $T_{aw}$ is the adiabatic wall temperature based on the bulk Mach number.

We choose the liner geometry to match as closely as possible the orifice size of acoustic liners in operating conditions. 
The acoustic liner comprises a total of 64 cavities: an array of $8 \times 4$ in the streamwise and spanwise direction on the upper and lower wall. 
Each cavity has a square cross-section with a side length $\lambda = 0.335 \delta$, 
depth $h = 0.5 \delta$. The orifices have a diameter of $d = 0.08 \delta$, 
the cavity walls have a thickness of $0.2 \delta$. 

We carry out simulations at three friction Reynolds numbers $Re_\tau = {500, 1000, 2000}$, corresponding to a viscous-scaled diameters $d^+ = {40, 80, 160}$. 
Additionally, we increase the liner porosity between $\sigma \approx 0.036$--$0.32$ by varying the number of orifices per cavity between 1 and 9. 
We also vary the facesheet thickness. 
Cases $L_n$ have a facesheet thickness of $t = d$ and cases $L_{tn}$ have a facesheet thickness of $t=d/2$.
Details of all flow cases are reported in table~\ref{tab:cases}. 
The orifice configurations within a cavity, along with a sketch of the entire domain, are shown in figure \ref{fig:instant}. 
We compare the results of the liner simulations with smooth-wall simulations at approximately matching friction Reynolds numbers. 
Quantities that are non-dimensionalised by $\delta_v$ and $u_{\tau}$ are denoted by the `$+$' superscript.

The near wall flow is spatially inhomogeneous due to the acoustic liner. 
Therefore, flow statistics are calculated by averaging in time and over
the cavity phase $\lambda$ in the streamwise and spanwise direction, both using Favre ( $\widetilde{\cdot}$ ) and Reynolds ( $\overline{\cdot}$ ) ensemble averages,

\begin{equation}
f(x,y,z,t) = \widetilde{f}(x,y,z) + f''(x,y,z,t), \quad f(x,y,z,t) = \overline{f}(x,y,z) + f'(x,y,z,t).
\end{equation}

Additionally, we use angle brackets $\langle \cdot\rangle$ to
denote intrinsic averages (average over the fluid only) in the wall-parallel directions.

\begin{figure}
    \centering
        \includegraphics[scale = 0.07] {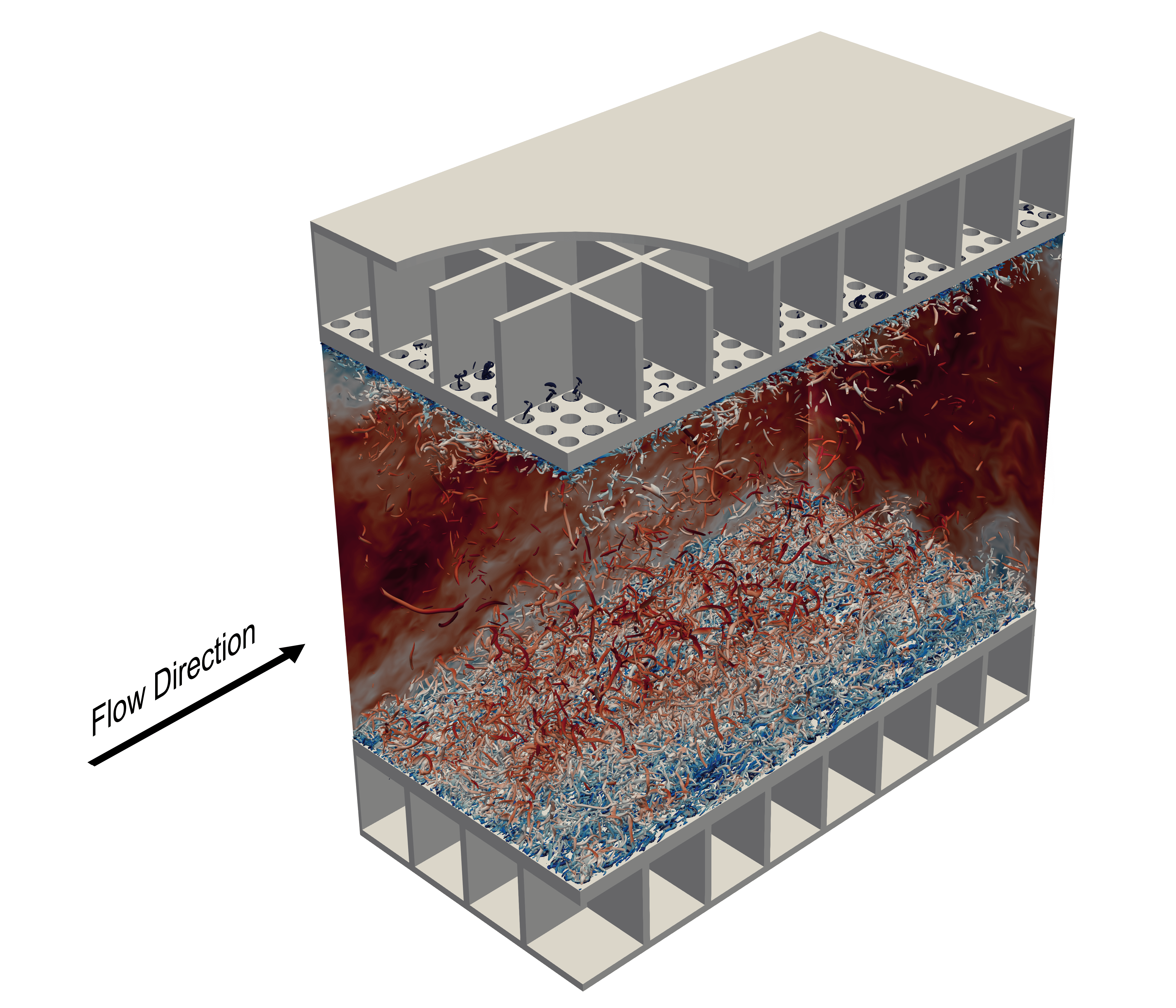} 
    \caption{Instantaneous flow field from DNS of turbulent channel flow at $Re_{\tau} = 2000$ and bulk Mach number $M_b = 0.3$. The streamwise velocity is shown in a $x-y$ plane, and in a $y-z$ plane. Vortical structures are visualised using the Q-Criterion.}
    \label{fig:instantboth}
\end{figure}

\section{Results}

\subsection{Instantaneous Flow}
\label{sec:inst}

\begin{figure}
    \centering
        \includegraphics[scale = 1] {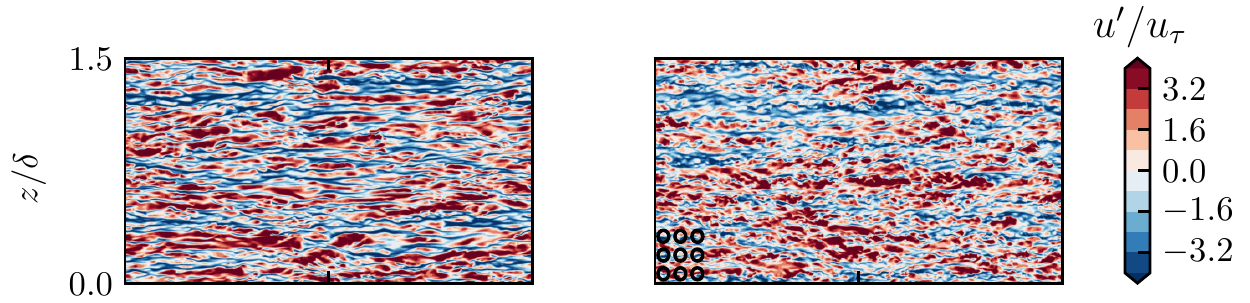}
        \put(-345,85){(\textit{a})}
        \put(-190,85){(\textit{b})}\\
        \includegraphics[scale = 1] {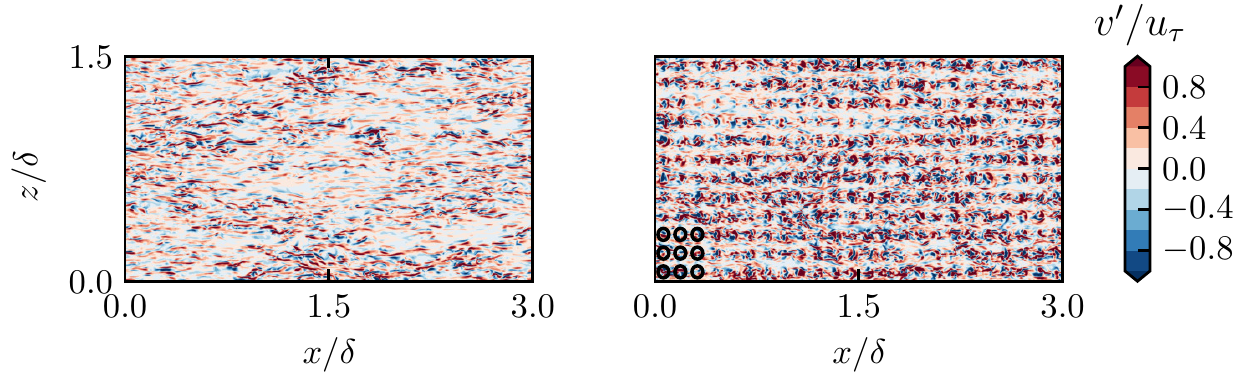} 
        \put(-345,100){(\textit{c})}
        \put(-190,100){(\textit{d})}\\
    \caption{Instantaneous streamwise velocity fluctuations in a wall-parallel plane at $y^+=12$ for flow case $S_3$ (\textit{left}) and flow case $L_6$ (\textit{right}) at $Re_{\tau} \approx 2000$. 
	On the figures for flow case $L_6$, the position of the orifices is shown at the bottom left corner, for one cavity.}
    \label{fig:fluc_12}
\end{figure}

\begin{figure}
    \centering
        \includegraphics[scale = 1] {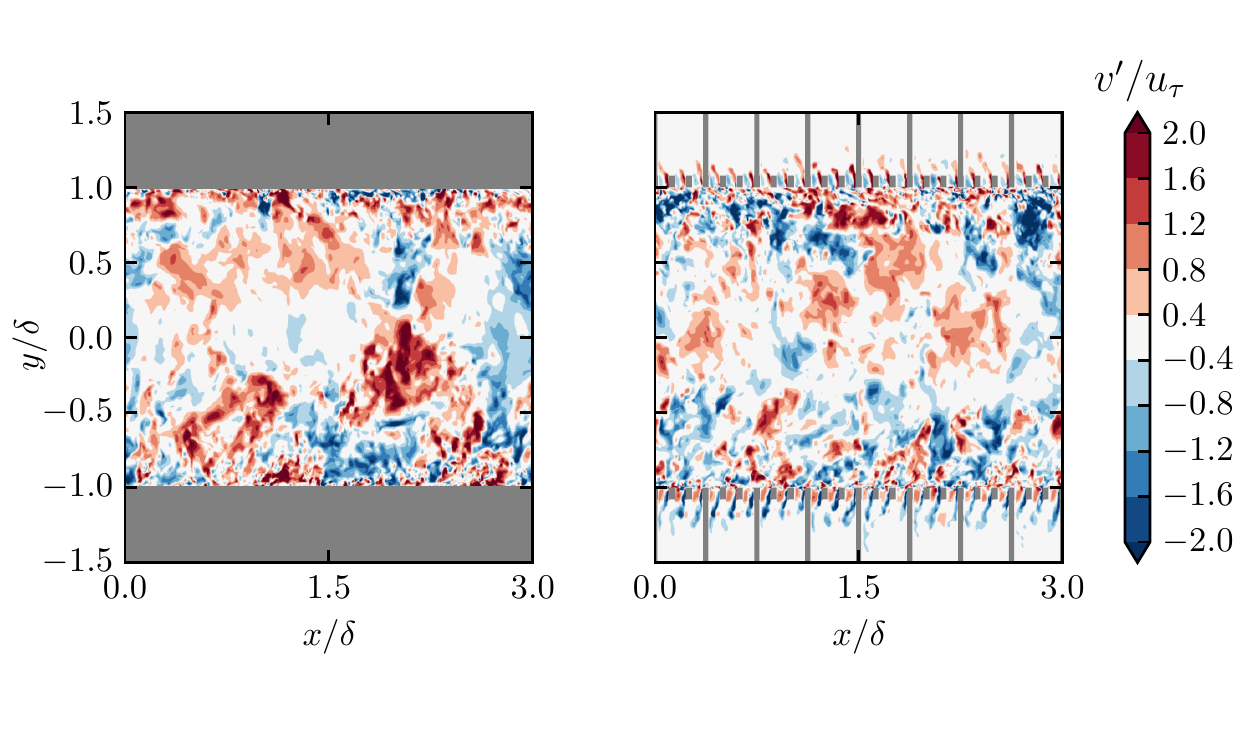} 
        \put(-345,200){(\textit{a})}
        \put(-190,200){(\textit{b})}\\
    \caption{Wall-normal velocity fluctuations in an $x-y$ plane at for flow case $S_3$ (\textit{a}) and flow case $L_5$ (\textit{b}) at $Re_{\tau} \approx 2000$. Grey patches represent solid wall regions.}
    \label{fig:fluc_xy}
\end{figure}

We begin our analysis by inspecting an instantaneous visualisation of flow case $L_6$
at friction Reynolds number $\Rey_\tau=2000$.
Figure \ref{fig:instantboth} shows the streamwise velocity in the wall-normal planes
and vortical structures visualised using the Q-Criterion. 
The near wall region is populated by small-scale structures indicating intense turbulence activity close to the wall,
whereas the flow below the cavities is more quiescent, although some vortices penetrate below the facesheet.

Figure \ref{fig:fluc_12} shows contours of the instantaneous streamwise (\textit{a}, \textit{b}) and wall-normal (\textit{c}, \textit{d}) velocity, in a wall-parallel plane above the facesheet for flow case $L_6$ taken $12$ wall units from the wall. 
The streamwise velocity is significantly altered as compared to the smooth wall
and near-wall streaks are shorter over the liner. 
The streaky structures can still be discerned, suggesting a modification, rather than a complete replacement of the near wall cycle. 

These observations are in line with previous studies on permeable walls,
which reported shorter streaks caused by the higher
wall-normal velocity fluctuations~\citep{kuwata_extensive_2019}. 
We also observe higher wall-normal velocity fluctuations as compared to the smooth wall, mainly concentrated
at the orifices, figure~\ref{fig:fluc_12}. 
The wall-normal velocity fluctuations seem reminiscent of
the underlying surface pattern, as the position of the
orifices can easily be discerned in the contours of $v'$, 
suggesting that turbulence in the
near-wall region is modulated by the surface topography~\citep{abderrahaman-elena_modulation_2019}.

Figure~\ref{fig:fluc_xy} shows a snapshot of the wall-normal velocity in an $x-y$ plane,
where we observe that
the effect of the liner on the flow is concentrated near the wall and inside the cavities. 
Inside the orifices, high wall-normal velocity fluctuations are visible, 
and they are notably higher at the downstream edge.
Wall-normal velocity fluctuations penetrate inside the cavities
forming a jet-like flow which extends down to $0.2\delta$ below the
facesheet, indicating important inertial effects inside the orifices.

\subsection{Mean Flow}

\begin{figure}
	\centering
	\includegraphics[scale = 1] {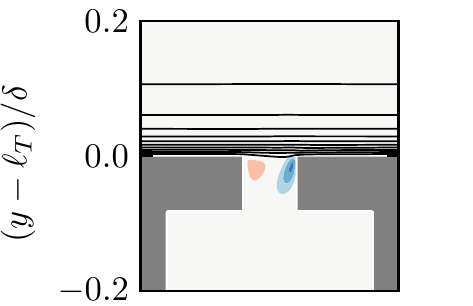}
	\put(-135,90){(\textit{a})}
	\includegraphics[scale = 1] {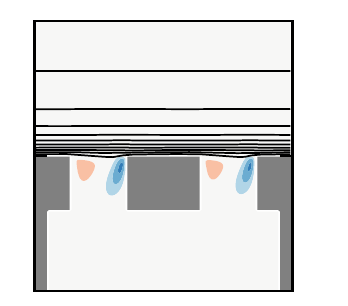} \put(-105,90){(\textit{b})}
	\includegraphics[scale = 1] {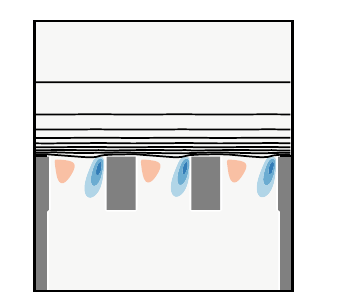}
	\put(-100,90){(\textit{c})} 
	\includegraphics[scale = 1] {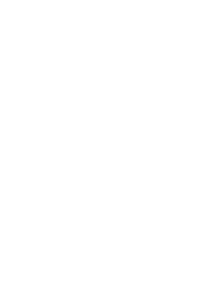}
	\\
	\includegraphics[scale = 1] {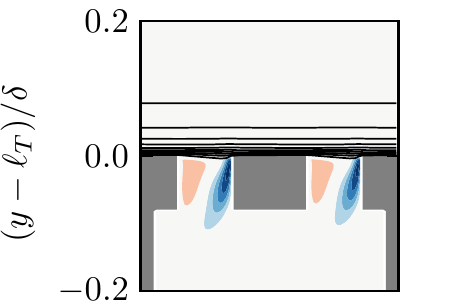}
	\put(-135,90){(\textit{d})}
	\includegraphics[scale = 1] {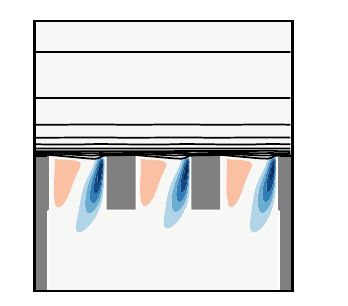} \put(-105,90){(\textit{e})}
	\includegraphics[scale = 1] {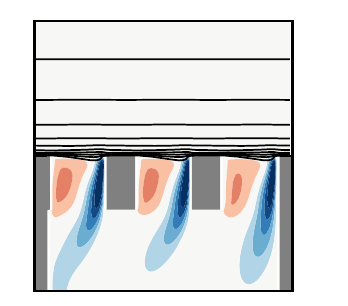}
	\put(-100,90){(\textit{f})} 
	\includegraphics[scale = 1] {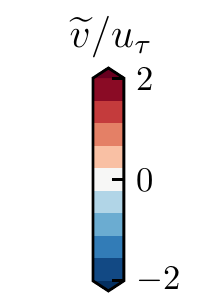}
	\\
	\includegraphics[scale = 1] {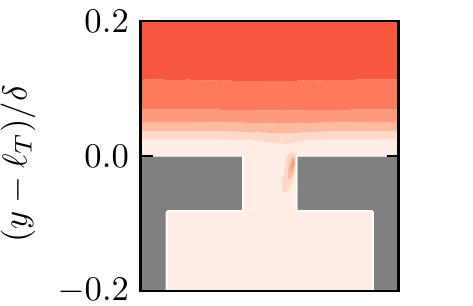}
	\put(-135,90){(\textit{g})}
	\includegraphics[scale = 1] {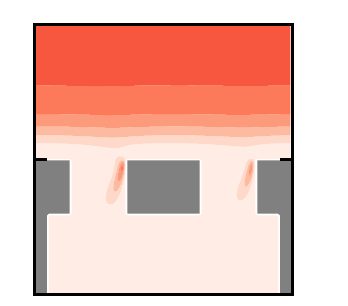} \put(-105,90){(\textit{h})}
	\includegraphics[scale = 1] {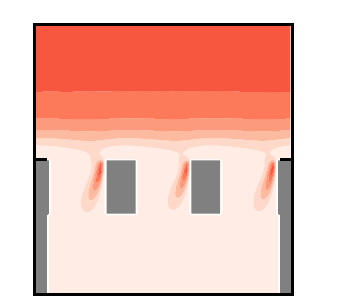}
	\put(-100,90){(\textit{i})} 
	\includegraphics[scale = 1] {PA_V2/colorbare-eps-converted-to.pdf}
	\\
	\includegraphics[scale = 1] {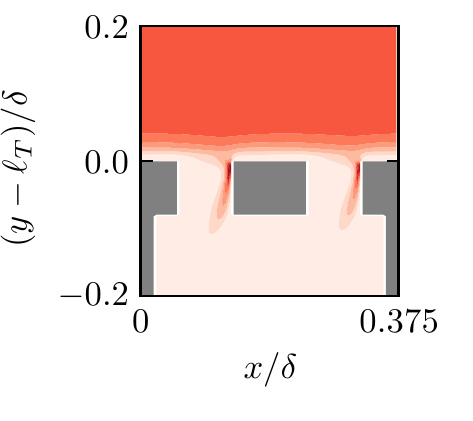}
	\put(-135,120){(\textit{j})}
	\includegraphics[scale = 1] {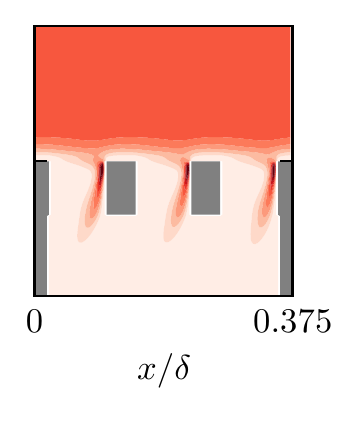} \put(-105,120){(\textit{k})}
	\includegraphics[scale = 1] {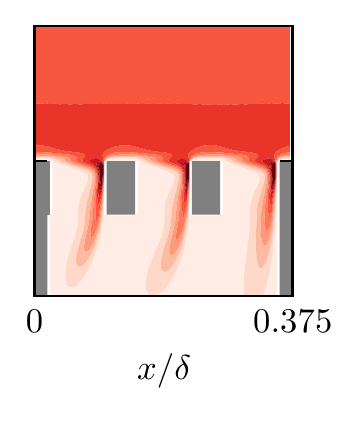}
	\put(-100,120){(\textit{l})} 
	\includegraphics[scale = 1] {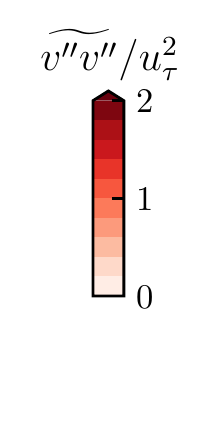}
	\\ 
	\caption{Mean wall-normal velocity $\widetilde{v}$ (\textit{a})--(\textit{f}) 
 	         and wall-normal Reynolds stress component $\widetilde{v'' v''}$  
 	         (\textit{g})--(\textit{l}) over a liner cavity
 	          for flow cases $L_1$ (\textit{a}, \textit{g}), $L_2$ (\textit{b}, \textit{h}),
 	          $L_3$ (\textit{c}, \textit{i}), $L_4$ (\textit{d}, \textit{j}),
 	          $L_5$ (\textit{e}, \textit{k}) and $L_6$ (\textit{f}, \textit{l}). 
 	          The isolines of the streamwise velocity are also shown in (\textit{a})--(\textit{f}).}
	\label{fig:pa}
\end{figure}

\begin{figure}
	\centering
	\includegraphics[scale = 1] {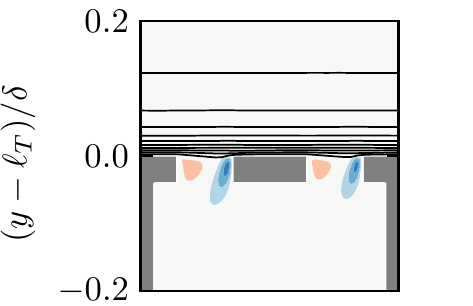}
	\put(-135,90){(\textit{a})}
	\includegraphics[scale = 1] {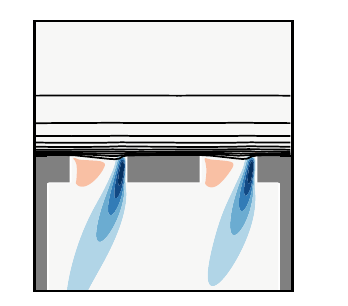} \put(-105,90){(\textit{b})}
	\includegraphics[scale = 1] {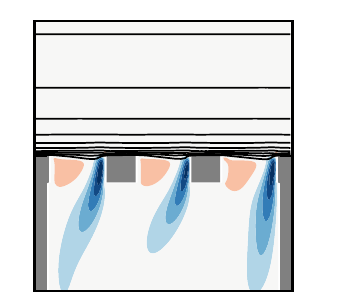}
	\put(-100,90){(\textit{c})} 
	\includegraphics[scale = 1] {PA_V2/colorbarv-eps-converted-to.pdf}
	\\
	\centering
	\includegraphics[scale = 1] {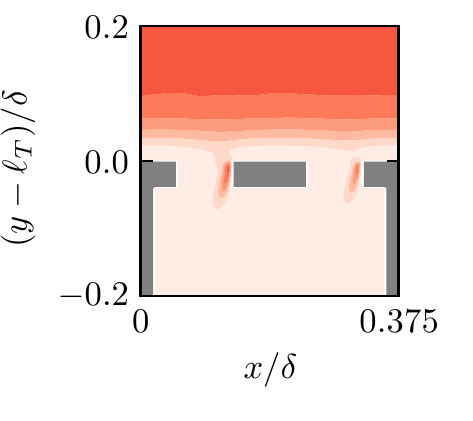}
	\put(-135,120){(\textit{d})}
	\includegraphics[scale = 1] {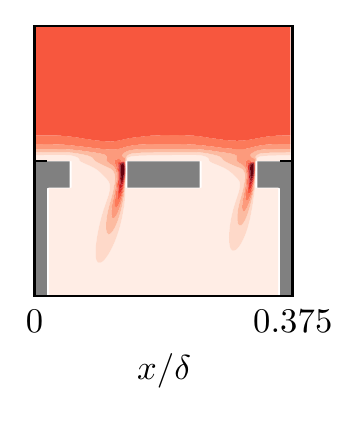} \put(-105,120){(\textit{e})}
	\includegraphics[scale = 1] {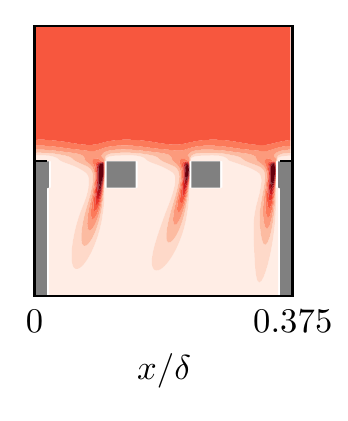}
	\put(-100,120){(\textit{f})} 
	\includegraphics[scale = 1] {PA_V2/colorbarvv-eps-converted-to.pdf}
	\\
	\caption{Mean wall-normal velocity $\widetilde{v}$ (\textit{a})--(\textit{c}) 
 	         and wall-normal Reynolds stress component $\widetilde{v'' v''}$  
 	         (\textit{d})--(\textit{f}) over a liner cavity
 	          for flow cases $L_{t2}$ (\textit{a}, \textit{d}), $L_{t4}$ (\textit{b}, \textit{e}) and
 	          $L_{t5}$ (\textit{c}, \textit{f}). 
 	          The isolines of the streamwise velocity are also shown in (\textit{a})--(\textit{c}).}
	\label{fig:pa_lt}
\end{figure}

In order to quantify the flow penetration
and inertial effects inside the orifices, we report
the mean wall-normal velocity, wall-normal Reynolds stress component, and isolines of the streamwise velocity
for cases $L_1$--$L_6$ in figure~\ref{fig:pa}.
Away from the facesheet, the flow
is homogeneous in the wall-parallel
directions, 
indicating that the effect of the liner is primarily contained in the near-wall region.
The mean flow is highly three-dimensional close to the liner.
The isolines of the streamwise velocity bend inward
at the downstream edge of the orifice where
the wall-normal velocity is negative, suggesting that
the flow penetrates inside the orifices. 
The wall-normal velocity is negative at the downstream edge of the orifice and positive at the upstream edge
due to the mean flow recirculation inside the pore, separating the region above and below the facesheet. 
The vortex is asymmetric, and the negative
values of $\widetilde{v}$ are always higher than the positive ones.
Moreover, we note that
the intensity of $\widetilde{v}$ is, primarily, 
a function of the viscous-scaled orifice diameter, whereas it seems less dependent on the porosity of the plate. 
 
For sufficiently large $d^+$, we observe high values
of the wall-normal velocity extending down into the cavity, 
resembling a jet-like flow also observed in the instantaneous flow in figure~\ref{fig:fluc_xy}.
This jet-like mean flow is accompanied by high wall-normal velocity fluctuations inside the orifice, as shown
in figures~\ref{fig:pa} (\textit{g})--(\textit{l}). 
Also the wall-normal velocity fluctuations
$\widetilde{v''v''}$ are higher at 
the downstream edge of the orifice, where they reach values comparable to, or even higher than, the peak in the near wall cycle.
This is particularly true for liner cases $L_5$ and $L_6$ 
(figures \ref{fig:pa} (\textit{k}), (\textit{l})) where $\widetilde{v''v''}$ is higher below the facesheet than in the near wall cycle above the facesheet.
These high wall-normal velocity fluctuations, 
are a symptom of inertial effects inside the orifices. 
A comparison between flow cases $L_1$--$L_6$ indicates that $\widetilde{v''v''}$ seems to depend on both $\sigma$ and $d^+$.

We also investigate the effect of
the plate thickness, using 
flow cases $L_{t2}$, $L_{t4}$ and $L_{t5}$,
which have $t=0.5d$.
Reducing the thickness causes an increase
of the mean wall-normal velocity (compare figure \ref{fig:pa} (\textit{e}) to \ref{fig:pa_lt} (\textit{c}))
and its fluctuations (compare figure \ref{fig:pa} (\textit{k}) to \ref{fig:pa_lt} (\textit{f})) within the orifice.
Wall-normal velocity fluctuations have been correlated with drag increase over rough surfaces~\citep{orlandi_dns_2006} 
as they are indicative of momentum transfer between the 
the crest and the through in the case of roughness, 
and the regions above and below the facesheet for acoustic liners. 
Therefore, this qualitative analysis suggests 
that the added drag over acoustic liners might depend on $d^+$, $\sigma$ and $t/d$, as we discuss further in the following section.

\subsection{Virtual Origin and Drag Increase}
\label{sec:drag}

\begin{figure}
    \centering
        \includegraphics[scale = 1] {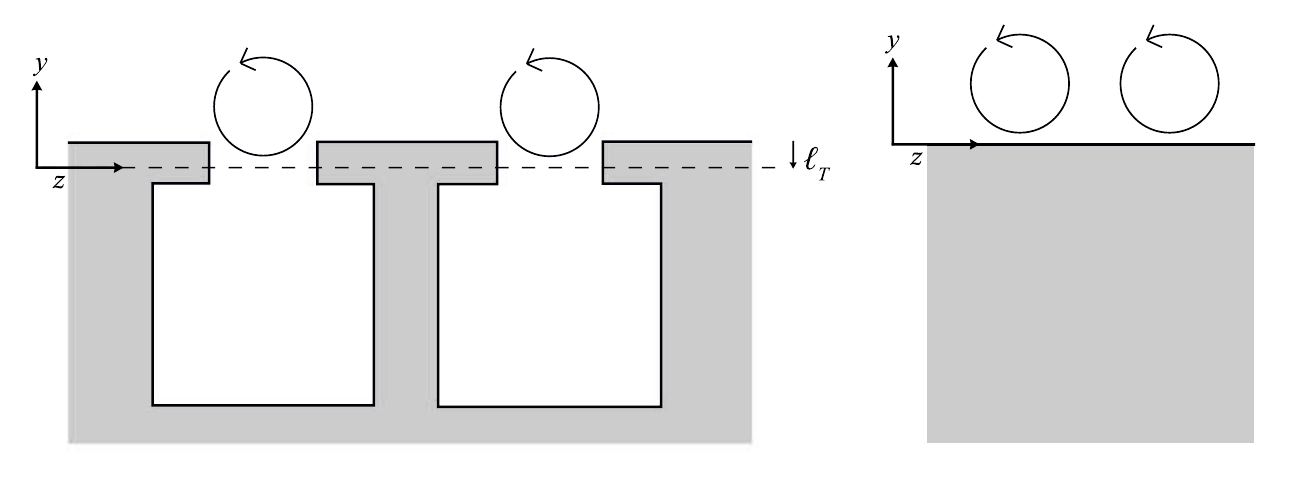}
	\caption{Schematic depicting the virtual origin of the flow configuration.}
    \label{fig:schemvo}
\end{figure}

\begin{figure}
    \centering
        \includegraphics[scale = 1] {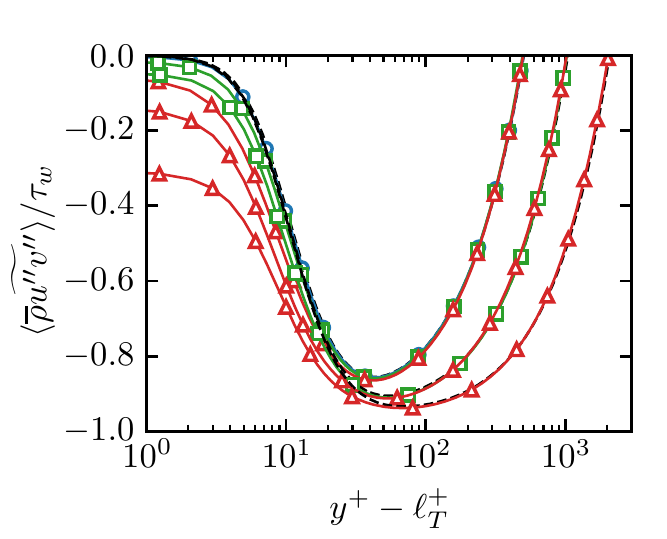}
        \put(-185,145){(\textit{a})}
        \includegraphics[scale = 1] {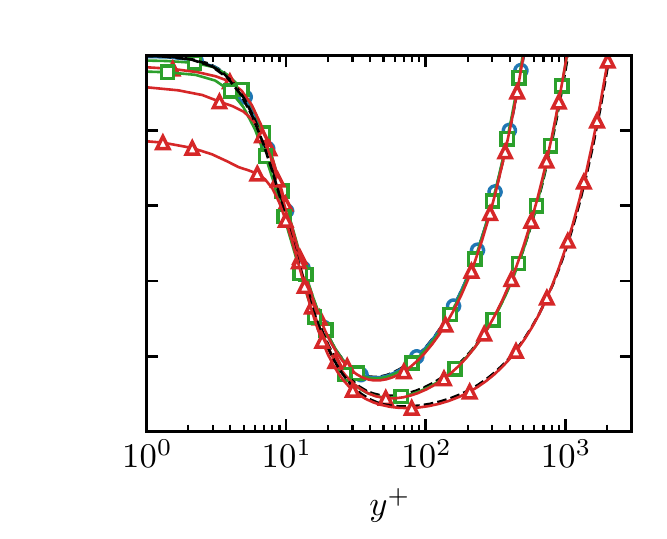}
        \put(-185,145){(\textit{b})} \\
    \caption{Intrinsic averaged Reynolds shear stress $\tau_{12}$ as a function of the wall-normal distance for smooth wall flow cases $S_1$ -- $S_3$ (dashed) and liner flow case $L_1$ -- $L_5$ (solid with symbols),
    before virtual origin correction
    (\textit{a}) and  
    after virtual origin correction (\textit{b}).
    (\textit{a}) Symbols indicate different porosities: $\sigma=0.0357$ (circles), $\sigma=0.143$ (squares) and $\sigma=0.322$ (triangles).}
    \label{fig:mean_uv}
\end{figure}

\begin{figure}
    \centering
        \includegraphics[scale = 1] {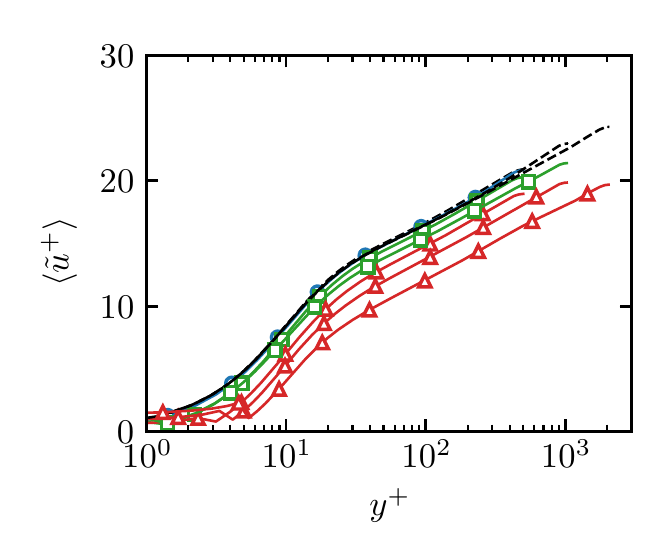}
        \put(-185,145){(\textit{a})}
        \includegraphics[scale = 1] {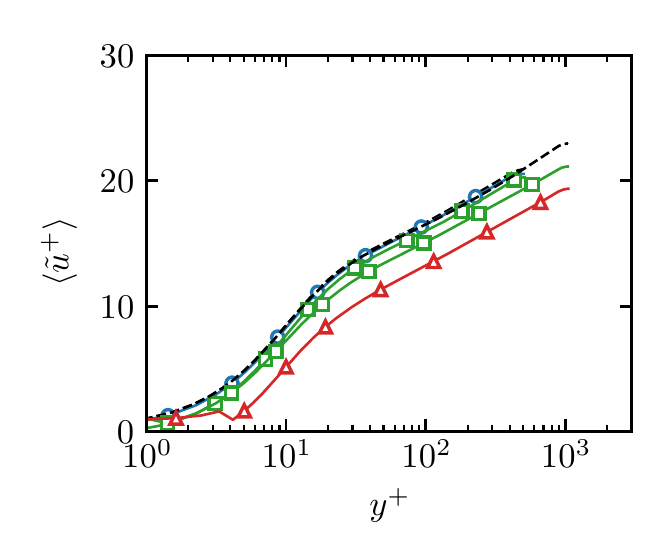}
        \put(-185,145){(\textit{b})} \\
    \caption{Intrinsic averaged mean streamwise velocity for for smooth wall flow cases $S_1$ -- $S_3$ (Dashed lines) and liner flow cases $L_n$ (\textit{a}) and $L_{tn}$ (\textit{b}) as a function of the wall-normal distance. Symbols indicate different porosities: $\sigma=0.0357$ (circles), $\sigma=0.143$ (squares) and $\sigma=0.322$ (triangles).}
    \label{fig:mean_vel}
\end{figure}

On smooth walls, there is no ambiguity
on the wall-normal origin of the flow, which is always at the wall, where both the mean velocity and Reynolds stresses are zero.
The presence of complex surface patterns 
introduces an uncertainty on the wall-normal origin location, 
which can be relevant when comparing rough wall results to the solution for a corresponding smooth wall.

This virtual origin is a flow property, and it can be interpreted as the wall-normal location
where the outer flow perceives the wall. 
Several methods to estimate the virtual origin have been proposed~\citep{jackson_displacement_1981, choi_direct_1993, modesti_dispersive_2021}. 
In the present work, we calculate the origin of turbulence $\ell_T$ following the 
approach of~\citet{ibrahim_smooth-wall-like_2021}, namely
we shift the Reynolds shear stress profile of the liner cases
to match the smooth wall one.
The virtual origin is located $\ell_T$
below the surface of the facesheet (figure~\ref{fig:schemvo}),
meaning that the near wall cycle tends to penetrate inside the orifices,
as is also clear from the high values of the Reynolds shear stress
in figure~\ref{fig:mean_uv} (\textit{a}), and from instantaneous flow visualizations.
The virtual origin shift is limited to a few wall units $\ell_T^+ < 4$ for all flow cases, but accounting for this displacement allows us to 
restore a very good match with the smooth wall data down
to the viscous sublayer (figure~\ref{fig:mean_uv} (\textit{b})), confirming that at least part of the effect
of the liner can be accounted for by an origin shift.

Having estimated the virtual origin, we can now draw meaningful comparisons between the smooth wall and liner statistics. 
Figure \ref{fig:mean_vel} shows the mean velocity profiles in viscous units for all flow cases. 
The mean velocity profiles over the liner show a downward shift $\Delta U^+$ with the respect
to the baseline smooth wall, indicating that the flow experiences higher drag.
Despite the shift, velocity profiles are parallel to each other, which supports outer-layer similarity, as typical of many rough surfaces~\citep{chung_predicting_2021}.
The von K\'arm\'an constant is $\kappa\approx 0.39$ for both liner and smooth wall cases.

This is in contrast to the work of \citet{breugem_influence_2006} and \citet{kuwata_transport_2016}, 
who reported different values of $\kappa$ over permeable surfaces. The discrepancy could be due to the low-Reynolds number of previous studies (maximum $Re_\tau \approx 350$ for smooth impermeable cases), or perhaps
to the use of Darcy-type boundary conditions, as compared
to pore-resolved simulations.
The flow cases with low porosity, $\sigma=0.0357$ and $d^+=40$ (circles), show a smooth-wall-like behaviour with very minor changes in the mean velocity profile. 
However, a departure from the smooth-wall velocity profile becomes evident as either $\sigma$ or $d^+$ is increased or $t/d$ is decreased. 

\begin{figure}
    \centering
        \includegraphics[scale = 1] {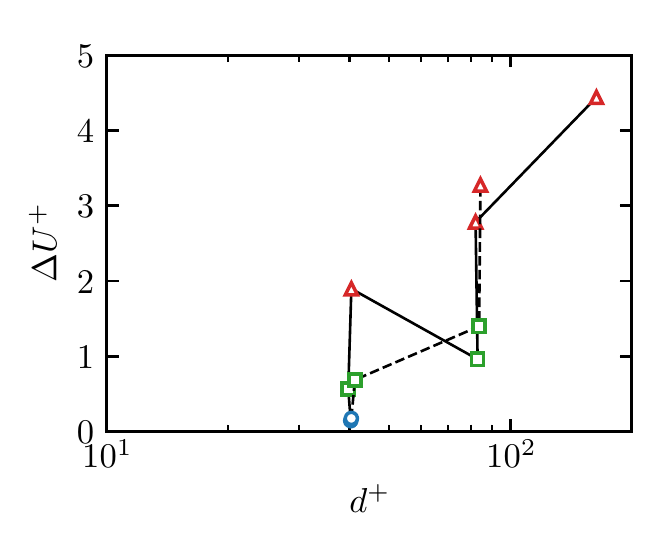}
        \put(-185,145){(\textit{a})}
        \includegraphics[scale = 1] {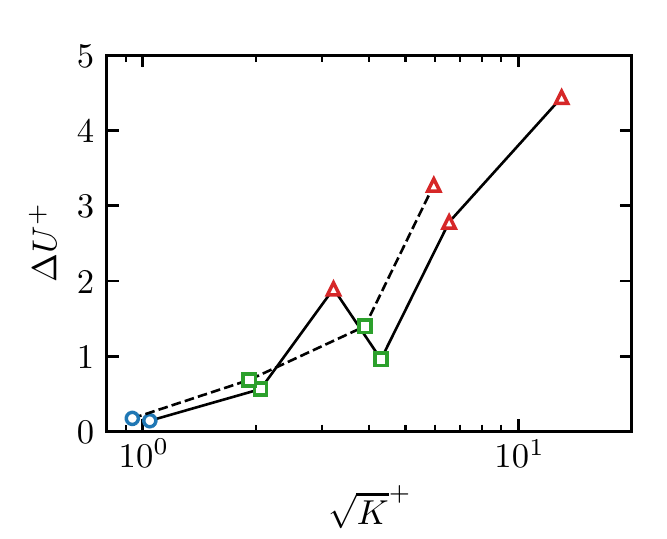}
        \put(-185,145){(\textit{b})} \\
    \caption{$\Delta U^+$ as a function of the viscous-scaled orifice diameter, $d^+$ (\textit{a}) and the Darcy permeability (\textit{b}). Different line types indicate different facesheet thickness: solid ($t=d$) and dashed ($t=d/2$).}
    \label{fig:deltau_wrong}
\end{figure}

\begin{figure}
    \centering
        \includegraphics[scale = 1] {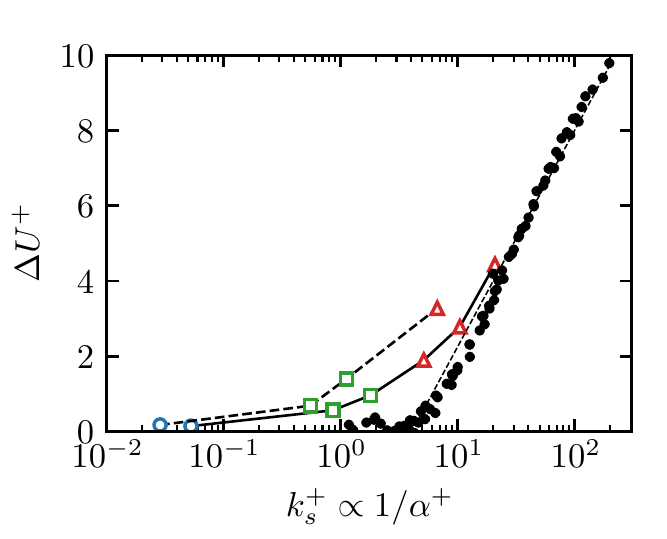}
        \put(-185,145){(\textit{a})}
        \includegraphics[scale = 1] {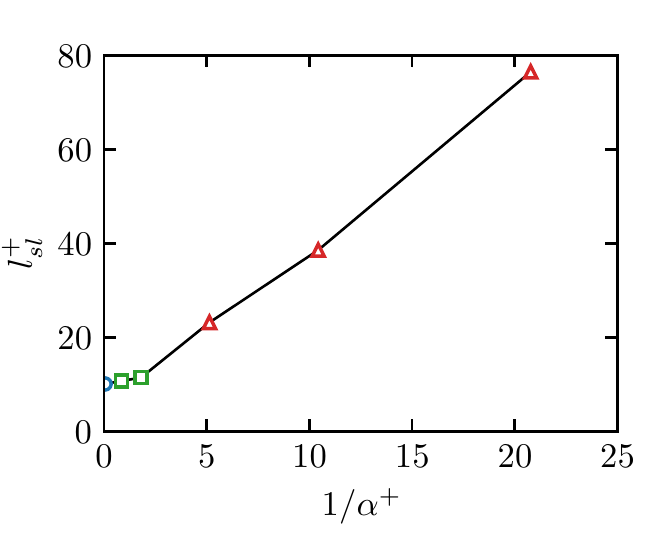}
        \put(-185,145){(\textit{b})} \\
    \caption{$\Delta U^+$ (\textit{a}) and the roughness sublayer (\textit{b}) as a function of the inverse of the Forchheimer coefficient, $1/\alpha^+$. Different line types indicate different facesheet thickness: Solid ($t=d$) and dashed ($t=d/2$). The thin dashed line in (\textit{a}) indicates $\Delta U^+ = \kappa^{-1} \text{log}(1/\alpha^+)-3.5$. The black filled circles indicate Nikuradse's data.}
    \label{fig:deltau}
\end{figure}

A fundamental question is whether acoustic liners
exhibit a fully rough regime, namely, whether the Hama roughness function follows a logarithmic law,

\begin{equation}
\Delta U^+ = \frac{1}{\kappa} \log(\ell^+ ) + B(\ell^+),
\end{equation}

where $\ell$ is a suitable length scale of the liner geometry. 
In canonical $k$-type roughness, $\ell$ is simply the roughness height, however, for acoustic liners different choices are possible. 
Unlike canonical roughness, there is no protrusion into the flow and therefore the definition of a suitable length scale is not straightforward.
It is clear that it depends upon the geometrical parameters of the orifice, 
namely the porosity, orifice diameter and plate thickness.
However, as is apparent in figure \ref{fig:deltau_wrong} (\textit{a}), none of these parameters can account
for the effect of the liner on their own. 
For instance, flow cases $L_2$ and $L_3$
have the same $t/d$ and approximately the same $d^+$, but different porosity and therefore a different $\Delta U^+$.
Similarly, cases $L_3$ and $L_5$ have the same porosity and $t/d$, but case $L_5$ has a larger viscous scaled diameter, leading to a larger $\Delta U^+$, see table~\ref{tab:cases}.
An increase in $\Delta U^+$ is also noted if the thickness is decreased and the other two parameters are constant.

Other candidate length scales can be inferred by regarding acoustic liners as porous
surfaces. The flow normal to a porous plate is characterized by the pressure drop through
the facesheet $\Delta P$, which can be expressed as the sum of two contributions~\citep{bae_numerical_2016, lee_modeling_1997},

\begin{equation}
    \dfrac{\Delta P}{t} \, \dfrac{d^2}{\rho\nu U_t} = \dfrac{d^2}{K} + \sigma \alpha d Re_p,
\end{equation}

where $\Rey_p = d U_p/\nu$ is the pore Reynolds number, 
$U_p$ is the volume averaged wall-normal velocity inside the orifice, 
$U_t = \sigma U_p$ is the superficial velocity, 
$K$ is the Darcy permeability coefficient, and $\alpha$ is the Forchheimer coefficient. 

The Darcy permeability has the physical dimension of an area whereas the Forchheimer coefficient
is the inverse of a length scale, and they are both related to the ease with which the flow passes through the plate because both contribute to the pressure drop.
Their relative importance depends on $\Rey_p$:
Darcy permeability dominates at low pore Reynolds number ($\Rey_p \leq\mathcal{O}(1)$), whereas the Forchheimer permeability becomes relevant from $Re_p\gtrsim5$. 
An extensive discussion on Darcy and Forchheimer drag is available in our recent work~\citep{shahzad_permeability_2022}, where we calculate the Darcy permeability and the Forchheimer coefficient
of perforated plates that match the present DNS dataset and compare the results to available engineering formulas. 

If we regard acoustic liners as porous surfaces, two relevant length scales
for the flow are the square root of the Darcy permeability $\sqrt{K}$ and the inverse of the Forchheimer coefficient $1/\alpha$, besides the orifice diameter. 
We show $\Delta U^+$ as a function of the viscous-scaled orifice diameter and the square root of the wall-normal Darcy permeability in figure \ref{fig:deltau_wrong}. 
As we noticed previously, the orifice diameter is not the proper length scale 
to characterize the added drag, and $\Delta U^+$ shows a non-monotonic trend when the porosity 
increases for constant $d^+$.
DNS data show that the square root of the Darcy permeability is also 
not suitable for predicting the drag increase, as we find a clear non-monotonic trend with $\sqrt{K}^+$, see figure \ref{fig:deltau_wrong}.

Instead, we find that $\Delta U^+$ shows a very promising trend
when reported as a function of the inverse of the viscous-scaled Forchheimer coefficient, suggesting that 
$1/\alpha^+$ is the most appropriate length scale for characterising the additional drag, figure~\ref{fig:deltau}.

This is consistent with the importance of inertia due to the very high wall-normal velocity fluctuations experienced inside the orifice, as observed in figure \ref{fig:pa}.
Hence, the Darcy permeability, which is commonly associated with the pressure drop in the limit case of Stokes flow, is 
no longer the dominant term. 
This is further elaborated upon in Section \ref{subsection:perm_fluc}. 
Additional supportive evidence 
that $1/\alpha$
is the relevant length scale is 
provided by figure~\ref{fig:deltau} (\textit{b}), showing
a nearly linear relation between
the inverse of the Forchheimer coefficient and the roughness sublayer.
The roughness sublayer is defined as the wall-normal location, measured from the virtual origin, where the time-averaged flow
becomes homogeneous in the wall-parallel directions~\citep{chung_predicting_2021}. 
It is a measure of the wall-normal extension
of the liner influence, and has been correlated often with the relevant roughness length scale \citep{raupach_rough-wall_1991,chan_18, modesti_dispersive_2021}.

Moreover, 
the data in figure~\ref{fig:deltau} (\textit{a})
show good agreement with data for classical sand-grain roughness 
of \citet{nikuradse_stromungsgesetze_1933}, supporting
the emergence of a fully rough regime,

\begin{equation}
    \Delta U^+(1/\alpha^+) = \frac{1}{\kappa} \log{(1/\alpha^+)} + C. 
    \label{eq:fully_rough}
\end{equation}
with $C \approx -3.5$.

For $t/d=1$, our data match very well the sand-grain roughness of~\citet{nikuradse_stromungsgesetze_1933}
with $k_s^+ \approx 1/\alpha^+$ 
being the equivalent sand-grain roughness height.
For flow cases with a lower plate thickness $t/d=0.5$,
we observe a similar trend, although the fully rough regime 
is not reached, and flow cases at higher $1/\alpha^+$ would be necessary to determine $k_s^+$ more accurately.

The existence of a fully rough regime is in line with the observations of \citet{esteban_2022}, who note a fully rough regime in their experiments over porous foams.
Prior to the work of \citet{esteban_2022}, however, there was ambiguity regarding the existence of a fully rough regime for porous surfaces.
For instance, \citet{manes_11a} performed experiments
on permeable beds with gravel grains and found that the friction factor continued to increase with the Reynolds number. 
\cite{breugem_influence_2006} used a volume averaging approach to model packed beds, and also concluded that 
the fully rough regime may not exist for permeable walls.
Unlike the work of \citet{esteban_2022} who note that the Darcy permeability is the relevant parameter, 
the non-linear permeability appears to be more relevant for acoustic liners.

\begin{table}
	\centering
\begin{tabular}{lcccccc}
         & $1/\alpha^+$ & $C_f \times 10^3$ & $C_{f,v} \times 10^3$ & $C_{f,p} \times 10^3$ & $C_{f,v}/C_f(\%)$ & $C_{f,p}/C_f(\%)$  \\ 
\hline
$S_1$    & 0            & 4.578             & 4.578                 & 0                     & 100               & 0                  \\
$S_2$    & 0            & 3.791             & 3.791                 & 0                     & 100               & 0                  \\
$S_3$    & 0            & 3.201             & 3.201                 & 0                     & 100               & 0                  \\
$L_1$    & 0.0528       & 4.598             & 4.492                 & 0.106                 & 97.7              & 2.3                \\
$L_2$    & 0.859        & 4.855             & 4.389                 & 0.466                 & 90.4              & 9.6                \\
$L_4$    & 1.73         & 4.527             & 3.988                 & 0.539                 & 88.1              & 11.9               \\
$L_3$    & 5.14         & 5.539             & 4.149                 & 1.390                 & 74.9              & 25.1               \\
$L_5$    & 10.4         & 5.082             & 3.608                 & 1.474                 & 71.0              & 29.0               \\
$L_6$    & 20.8         & 5.267             & 3.029                 & 2.238                 & 57.5              & 42.5               \\ 
\hline
$L_{t1}$ & 0.0287       & 4.738             & 4.620                 & 0.118                 & 97.5              & 2.5                \\
$L_{t2}$ & 0.552        & 4.856             & 4.382                 & 0.476                 & 90.2              & 9.8                \\
$L_{t4}$ & 1.12         & 4.475             & 3.889                 & 0.586                 & 86.9              & 13.1               \\
$L_{t5}$ & 6.69         & 5.317             & 3.637                 & 1.680                 & 68.4              & 31.6               \\
\hline
\end{tabular}
	\caption{Contribution of pressure and viscous drag to the skin-friction coefficient of acoustic liners.}
	\label{tab:viscpress}
\end{table}

The fully rough regime is usually associated with the dominance
of pressure drag over viscous drag, and the same appears to
hold for acoustic liners.
In table \ref{tab:viscpress} we report the skin-friction 
coefficient, decomposed into its viscous and pressure
contribution, which shows that pressure drag is nearly negligible
for flow case $L_1$, whereas it becomes comparable to viscous drag
for flow case $L_6$.
The same trend is also observed for cases with lower plate thickness.
Even though pressure drag is still not contributing less than 50\% for flow case $L_6$, we believe that the trend is rather clear 
and it supports the emergence of a fully rough regime for acoustic liners.

\begin{figure}
	\centering
	\includegraphics[scale = 1] {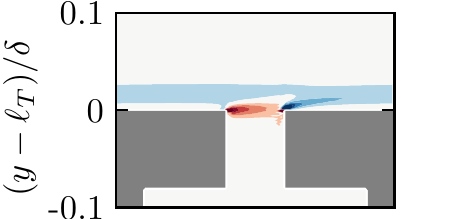}
	\put(-135,75){(\textit{a})}
	\includegraphics[scale = 1] {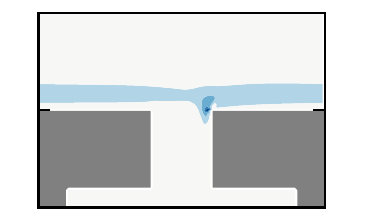}
	\put(-120,75){(\textit{b})}
	\includegraphics[scale = 1] {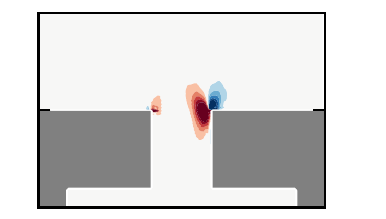}
	\put(-120,75){(\textit{c})} 
	\\
	\includegraphics[scale = 1] {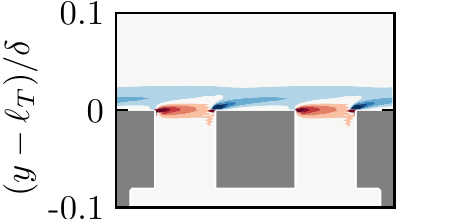}
	\put(-135,75){(\textit{d})}
	\includegraphics[scale = 1] {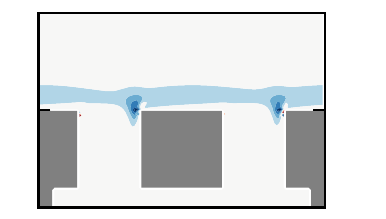}
	\put(-120,75){(\textit{e})}
	\includegraphics[scale = 1] {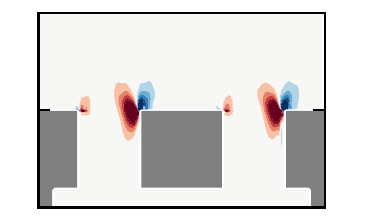}
	\put(-120,75){(\textit{f})} 
	\\
	\includegraphics[scale = 1] {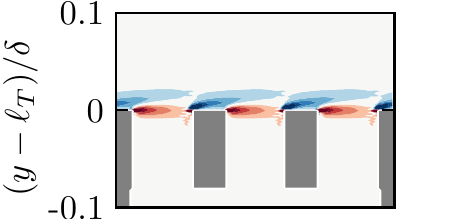}
	\put(-135,75){(\textit{g})}
	\includegraphics[scale = 1] {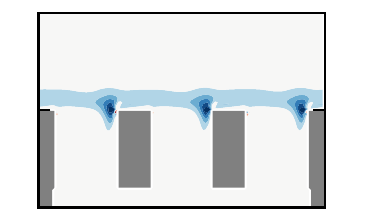}
	\put(-120,75){(\textit{h})}
	\includegraphics[scale = 1] {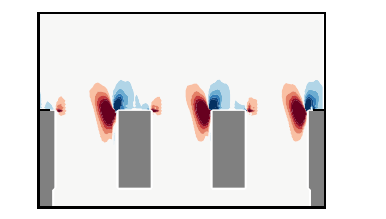}
	\put(-120,75){(\textit{i})} 
	\\
	\includegraphics[scale = 1] {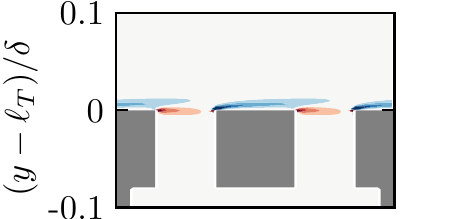}
	\put(-135,75){(\textit{j})}
	\includegraphics[scale = 1] {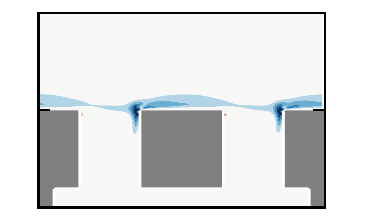}
	\put(-120,75){(\textit{k})}
	\includegraphics[scale = 1] {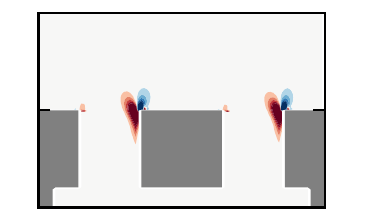}
	\put(-120,75){(\textit{l})} 
	\\	
	\includegraphics[scale = 1] {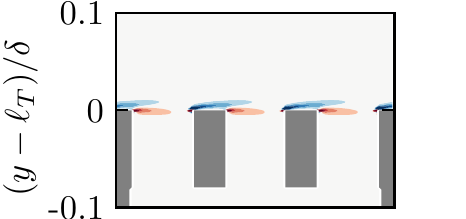}
	\put(-135,75){(\textit{m})}
	\includegraphics[scale = 1] {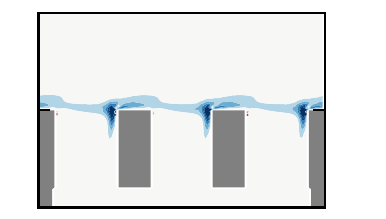}
	\put(-120,75){(\textit{n})}
	\includegraphics[scale = 1] {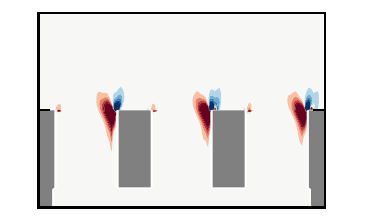}
	\put(-120,75){(\textit{o})} 
	\\
	\includegraphics[scale = 1] {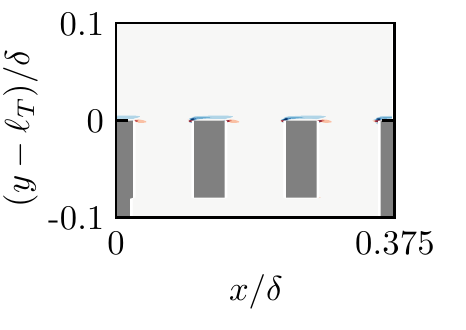}
	\put(-135,100){(\textit{p})}
	\includegraphics[scale = 1] {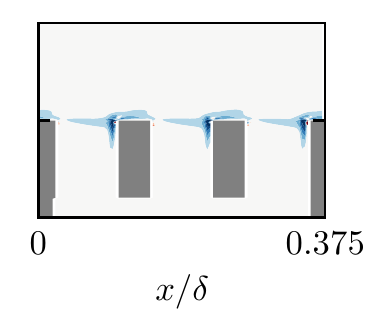}
	\put(-120,100){(\textit{q})}
	\includegraphics[scale = 1] {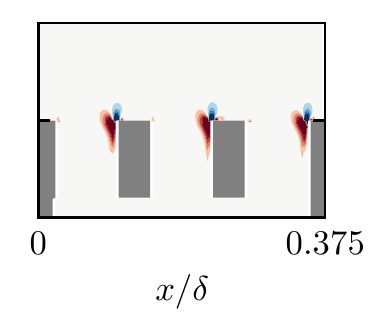}
	\put(-120,100){(\textit{r})} 
	\\
	\includegraphics[scale = 1] {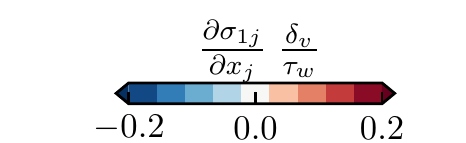}
	\includegraphics[scale = 1] {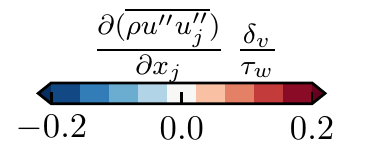}
	\includegraphics[scale = 1] {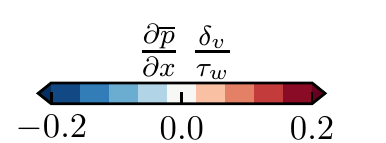} \\
	\caption{Contours of viscous diffusion (left),  turbulent convection (middle) and pressure gradient (right), normalised by $\delta_v/\tau_w$ for cases $L_1$--$L_6$ (top to bottom).}
	\label{fig:mombal}
\end{figure}

The relevance of pressure drag can also be demonstrated by analysing the mean momentum balance in the streamwise direction,

\begin{equation}
    \dfrac{\partial  \overline{\rho} \widetilde{u} \widetilde{u_j} }{\partial x_j}
  + \dfrac{\partial  \overline{\rho}\widetilde{u'' u_j''} }{\partial x_j} = 
  - \dfrac{\partial  \overline{p} }{\partial x} 
  + \dfrac{\partial  \widetilde{\sigma}_{1j} }{\partial x_j}
  + \overline{\Pi}.
  \label{eq:mmb}
\end{equation}

Figure \ref{fig:mombal} shows the contribution
of the different terms in equation~\eqref{eq:mmb}, 
close to the orifice.
Viscous diffusion becomes less relevant as the Reynolds
number increases, whereas the intensity of turbulent convection
increases, although the maximum value is confined very close to 
the wall, and inside the orifices.
The magnitude of the pressure gradient term is constant for all considered Reynolds number,
and its maximum location shifts downward into the orifices as the Reynolds number increases.
The figure shows that the contribution of the pressure gradient is significant and its relative importance grows
as the viscous sublayer becomes thinner.
We also note that increasing the number of holes (porosity)
increases the pressure drag, as each orifice seem
to contribute approximately the same, independently 
of its location.

The values of the friction coefficient reported in table~\ref{tab:viscpress}
only apply to the Reynolds number of the present DNS, which is much lower than in
a realistic configuration.
Fortunately, the existence of a fully rough regime simplifies the modelling
of acoustic liners and makes it easy to estimate the drag increase they induce in 
operating conditions. As discussed in Section~\ref{sec:introduction},
the friction Reynolds number over acoustic liners is $\Rey_\tau\approx6600$,
and the viscous-scaled inverse of the Forchhemeir coefficient
for a geometry with $d/\delta\approx0.08$, $t/d\approx1$ and $\sigma\approx 0.3$ is $1/\alpha^+\approx80$~\citep{shahzad_permeability_2022}.
Assuming that a fully rough regime exists, then equation~\eqref{eq:fully_rough}
returns $\Delta U^+\approx7$, which can be converted into drag variation~\citep{modesti_dispersive_2021},

\begin{equation}
\Delta \text{Drag}(\%) =  \frac{1}{\left(1-\frac{\Delta U^+}{u_{\infty,s}^+}\right)},
\end{equation}
where $u_{\infty,s}$ is the freestream velocity over the smooth wall.
Hence, acoustic liners are expected to provide about $70\%$ drag increase
per plane area with respect to a smooth wall. Of course, this value
might slightly change depending on the specific geometry considered, and the presence 
of incoming acoustic waves could also alter this result.
\subsection{Permeability and Velocity Fluctuations}
\label{subsection:perm_fluc}

\begin{figure}
	\centering
	\includegraphics[scale = 1,trim={0 0.75cm 0 0},clip] {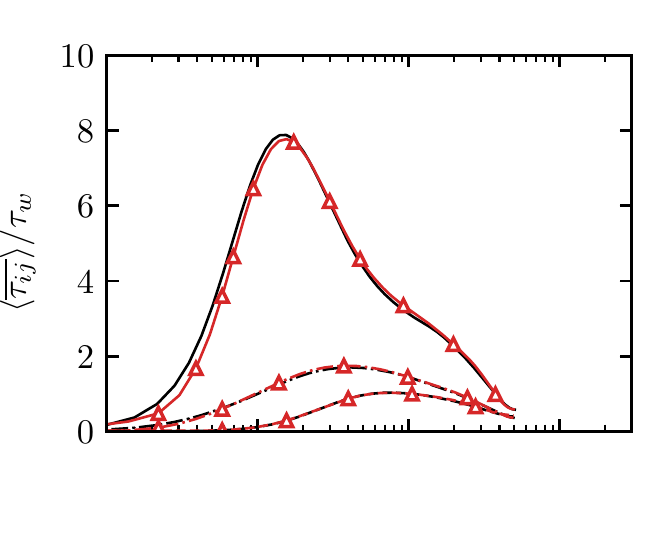}
	\put(-185,130){(\textit{a})}
	\includegraphics[scale = 1,trim={0.5cm 0.75cm 0 0},clip] {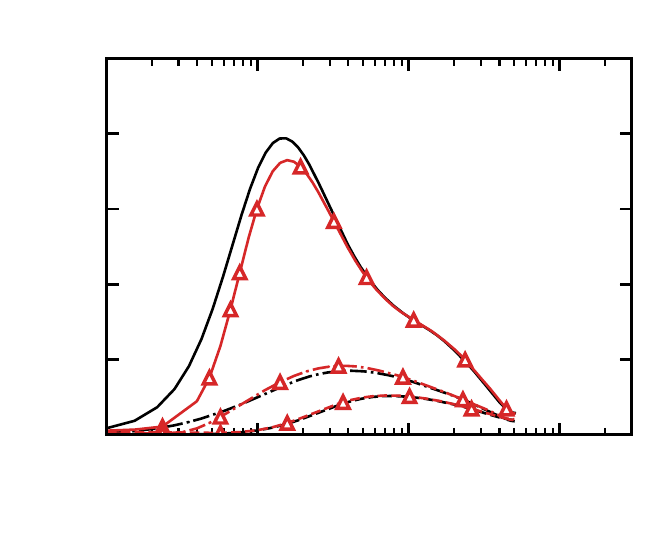}
	\put(-175,130){(\textit{b})} \\
	\includegraphics[scale = 1] {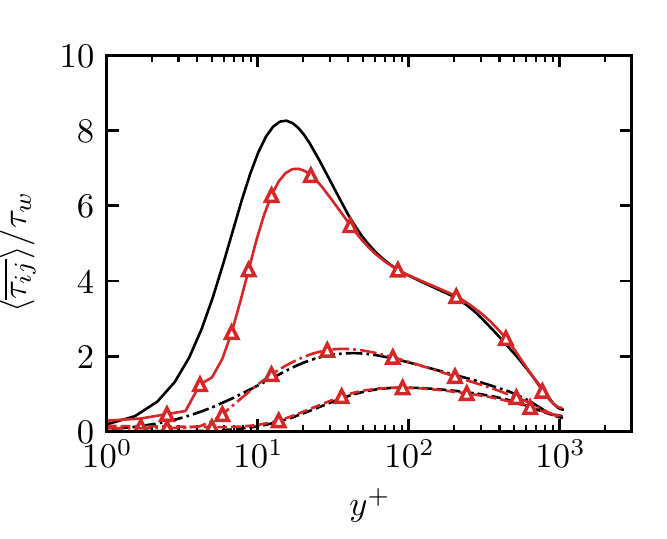}
	\put(-185,148){(\textit{c})}
	\includegraphics[scale = 1,trim={0.5cm 0 0 0},clip] {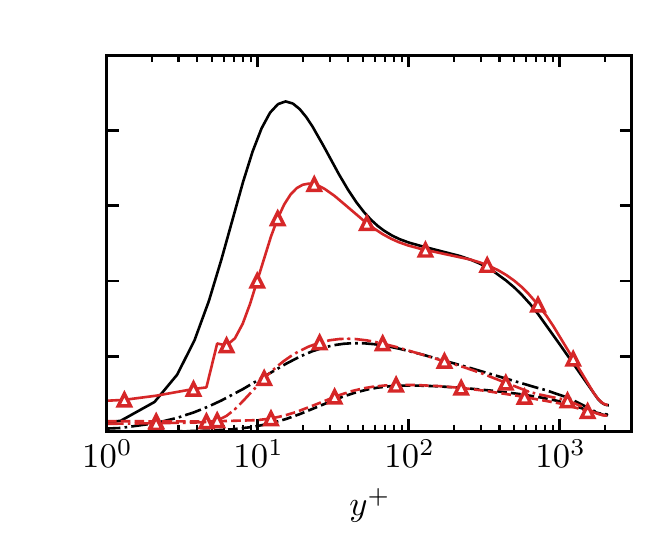}
	\put(-175,148){(\textit{d})}
	\caption{Intrinsic averaged Reynolds stresses as a function of the viscous-scaled wall-normal distance for flow case $L_2$ with $1/\alpha^+ = 0.859$ (\textit{a}), $L_3$ with $1/\alpha^+ = 5.14$ (\textit{b}), $L_5$ with $1/\alpha^+ = 10.4$ (\textit{c}) and flow case $L_6$ with $1/\alpha^+ = 20.8$ (\textit{d}). Lines without symbols indicate the smooth-wall cases and the triangles indicate the liner case. Solid lines indicate $\tau_{11}/\tau_w$, dashed lines indicate $\tau_{22}/\tau_w$ and dashed-dotted lines indicate $\tau_{33}/\tau_w$.}
	\label{fig:rs}
\end{figure}

\begin{figure}
	\centering
	\includegraphics[scale = 1] {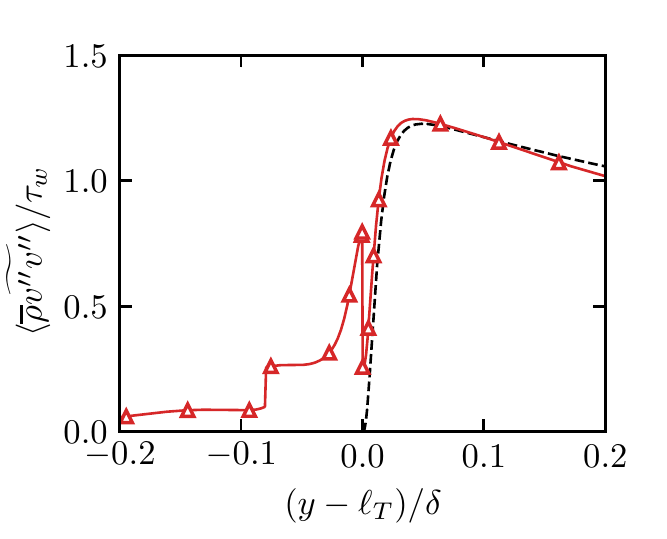}
	\put(-185,145){(\textit{a})}
	\includegraphics[scale = 1] {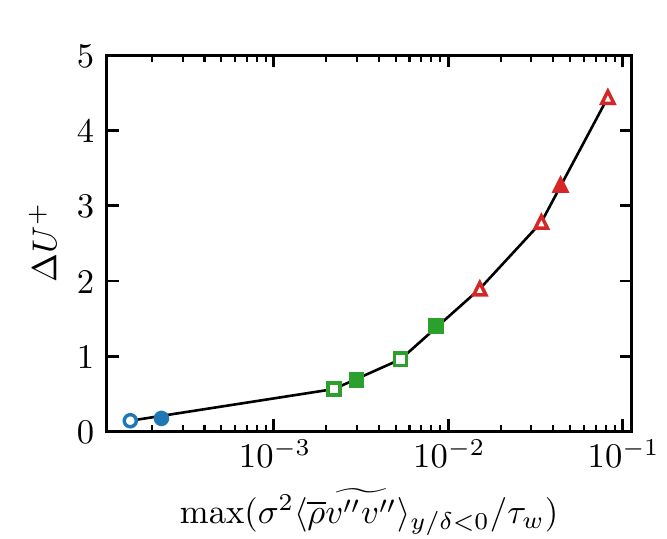}
	\put(-185,145){(\textit{b})}
	\caption{Intrinsic averaged wall-normal velocity fluctuations (\textit{a}) as a function of the wall-normal distance for flow case $L_6$ with $1/\alpha^+ = 20.8$ and $\Delta U^+$ (\textit{b}) as a function of maximum of the wall-normal velocity fluctuations below the wall. The dashed line in (\textit{a}) indicates the smooth-wall case. Different symbols indicate different facesheet thickness: Empty ($t=d$) and filled ($t=d/2$).}
	\label{fig:vrms}
\end{figure}

We further analyse the effect of acoustic liners on the Reynolds stresses, see figure \ref{fig:rs}.
Differences with respect to the smooth wall are primarily observed near the wall and increase as the permeability increases.
The relaxed impermeability condition gives rise to
non-zero Reynolds stresses at the liner wall,
thus enhancing momentum transfer between the flow above 
and below the plate.
The peak of the Reynolds stresses is also modified. 
The maximum of $\tau_{33}$ increases slightly, 
whereas the maximum of $\tau_{11}$ decreases, compared to the smooth wall, which has also been reported
for other types of porous surfaces~\citep{kuwata_lattice_2016,kuwata_extensive_2019}.

In the outer layer, the Reynolds stresses of the liner cases approximately match the smooth wall ones, as also typical of flows over rough walls.
Small differences in the outer layer are visible 
for cases $L_5$ and $L_6$, 
hinting at a possible departure from outer layer similarity as the viscous-scaled Forchheimer permeability decreases.
This is in contrast to what was observed for the mean streamwise velocity, whose outer layer seems to be more resilient to changes in the
underlying surface pattern.

Different authors noted a correlation
between wall-normal velocity fluctuations and drag variation
over roughness~\citep{orlandi_turbulent_2006,orlandi_dns_2006}, riblets~\citep{di_giorgio_relationship_2020},
and perforated plates~\citep{wilkinson_influence_1983}.
\citet{wilkinson_influence_1983} studied surfaces 
similar to acoustic liners geometries,
and proposed the increase of wall-normal velocity fluctuations as the root cause behind the
added drag. 
\citet{orlandi_turbulent_2006} and \citet{orlandi_dns_2006}
noted that the Hama roughness function of different 2D and 3D roughness geometries followed the same trend
when reported as a function of the wall-normal velocity fluctuations, suggesting a correlation
with the drag increase.
The present simulations appear to confirm this trend. 
First, we note that acoustic liners exhibit high wall-normal velocity fluctuations inside the orifice, as shown in figure \ref{fig:vrms} (\textit{a}) for flow case $L_6$.
The peak of $\langle\overline{\rho}\widetilde{v''v''}\rangle$ corresponds to the wall-normal location of
most intense wall-normal velocity fluctuations at the downstream edge in figure \ref{fig:pa}.

Figure~\ref{fig:vrms} shows the Hama roughness
function as a function of the maximum wall-normal Reynolds stress inside the liner orifice. 
Note that the latter has been weighted with the square of the porosity, 
following the idea that the reference velocity seen by the porous plate is the fluctuating superficial velocity, i.e. $\sigma v'$~\citep{bae_numerical_2016}.
All flow cases, irrespective of the thickness of the facesheet, follow the same trend,
pointing to a correlation
between velocity fluctuations and $\Delta U^+$,
as suggested by previous studies on rough surfaces.
Large values of the wall-normal 
velocity fluctuations inside the orifices
are clear symptoms of dominant inertial effects
in this region of the flow. 

This observation further supports the use of the Forchheimer permeability as the relevant length scale for the flow.
Although, there is no clear line demarcating where non-Darcy effects become dominant over the Darcy ones,
\citet{tanner_flowpressure_2019} note that already beyond pore Reynolds number $Re_p \approx 10$, it is necessary to account for non-Darcy effects. 
Using the maximum of the wall-normal velocity variance inside the orifice (weighted with $\sigma$), the pore Reynolds numbers for the present
flow cases are in the range $Re_p \approx 50-500$, 
which is well into the nonlinear regime of permeability.

To further investigate what these wall-normal velocity fluctuations arise from, we look at the budget of the wall-normal velocity variance,

\begin{equation}
      \dfrac{\partial}{\partial x_i} \left( \dfrac{1}{2}\widetilde{v'' v''} \Tilde{u_i} \overline{\rho} \right) = P_k + T_k + \Pi_k + \Pi_{\alpha \alpha}- \epsilon,
    \label{eq:vtke2}
\end{equation}

where

\begin{align*}
    P_k &= -\overline{\rho} \widetilde{v'' u_i''} \dfrac{\partial \Tilde{v}}{\partial x_i}, \;\;\;\; \epsilon = - \overline{\sigma_{i2} \dfrac{\partial v''}{\partial x_i}},\\
    T_k &= - \dfrac{\partial}{\partial x_i} \left(\dfrac{1}{2}\overline{\rho} \widetilde{ v'' v''  u_i''} - \overline{\sigma_{i2} v''}+\overline{p' v'}\delta_{2i}\right), \\
    \Pi_k &= -  \overline{\rho} \overline{v''} \dfrac{\partial \overline{p}}{\partial y}, \;\;\;\; \Pi_{\alpha \alpha} = \overline{p'\dfrac{\partial v'}{\partial y}}.
\end{align*}

\begin{figure}
    \centering
        \includegraphics[scale = 1] {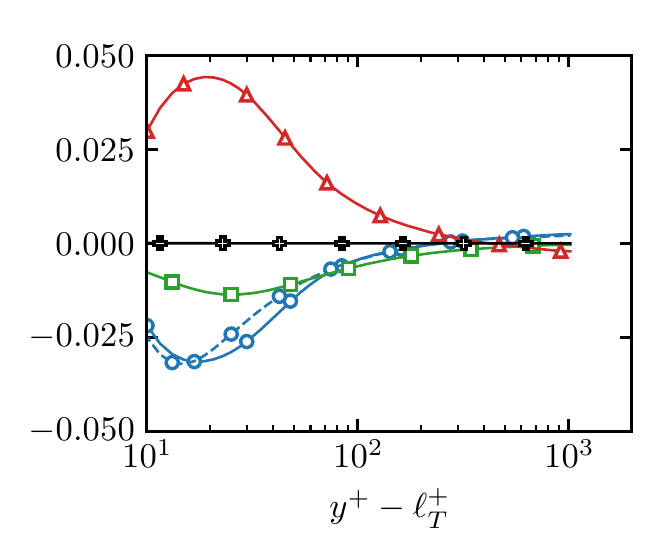}
        \put(-185,145){(\textit{a})}
        \includegraphics[scale = 1] {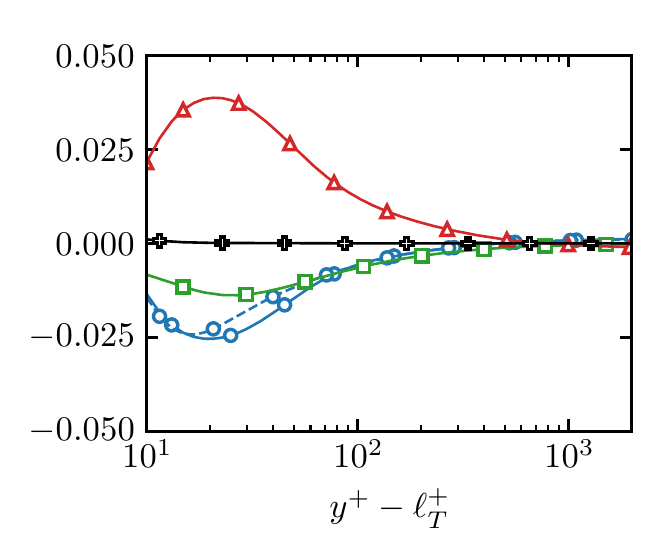}
        \put(-185,145){(\textit{b})} \\
    \caption{Intrinsically averaged turbulent wall-normal kinetic energy budget for flow case $L_5$ (\textit{a}) and $L_6$ (\textit{b}). Symbols represent different terms: $T_k$ (circles), $\epsilon$ (squares), $\Pi_{\alpha \alpha}$ (trianlges) and $P_k$ (plusses). The dashed line with circles represents the transport of wall-normal velocity fluctuations due to pressure fluctuations.}
    \label{fig:mean_vtke}
\end{figure}

\begin{figure}
	\centering
	\includegraphics[scale = 1] {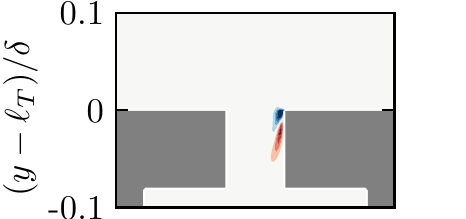}
	\put(-135,75){(\textit{a})}
	\includegraphics[scale = 1] {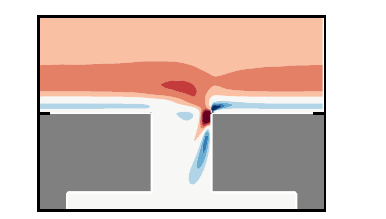}
	\put(-120,75){(\textit{b})}
	\includegraphics[scale = 1] {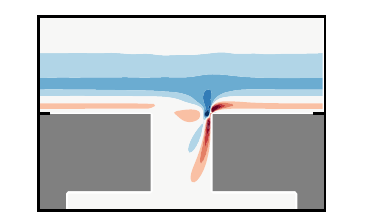}
	\put(-120,75){(\textit{c})} 
	\\
	\includegraphics[scale = 1] {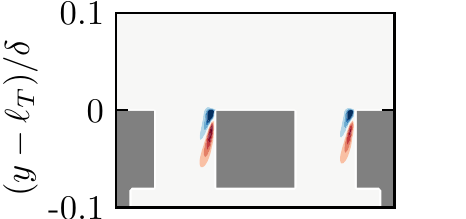}
	\put(-135,75){(\textit{d})}
	\includegraphics[scale = 1] {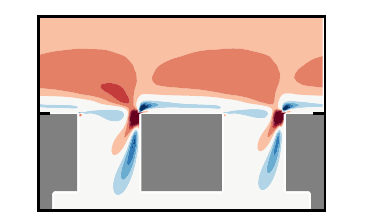}
	\put(-120,75){(\textit{e})}
	\includegraphics[scale = 1] {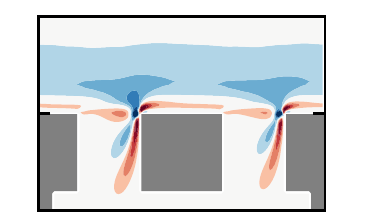}
	\put(-120,75){(\textit{f})} 
	\\
	\includegraphics[scale = 1] {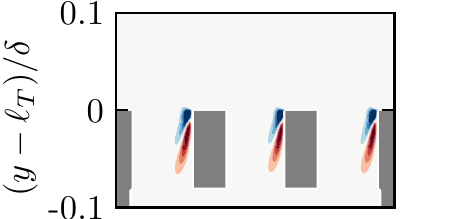}
	\put(-135,75){(\textit{g})}
	\includegraphics[scale = 1] {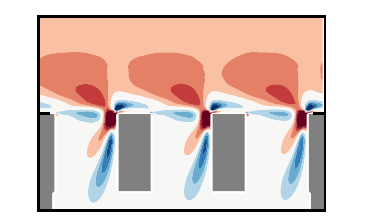}
	\put(-120,75){(\textit{h})}
	\includegraphics[scale = 1] {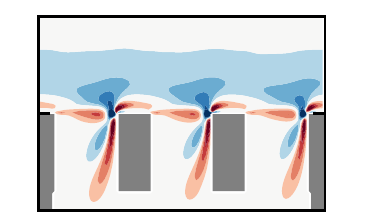}
	\put(-120,75){(\textit{i})} 
	\\
	\includegraphics[scale = 1] {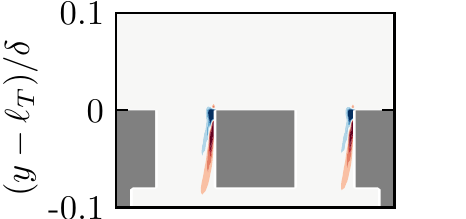}
	\put(-135,75){(\textit{j})}
	\includegraphics[scale = 1] {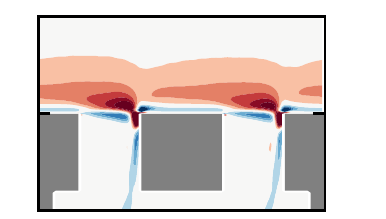}
	\put(-120,75){(\textit{k})}
	\includegraphics[scale = 1] {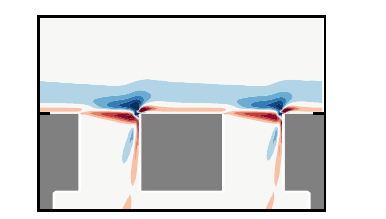}
	\put(-120,75){(\textit{l})} 
	\\	
	\includegraphics[scale = 1] {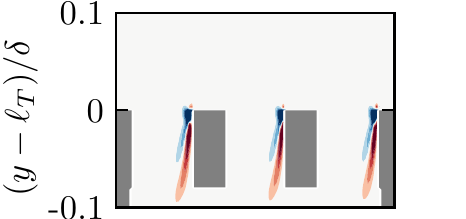}
	\put(-135,75){(\textit{m})}
	\includegraphics[scale = 1] {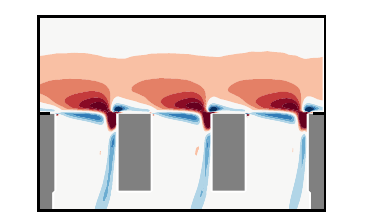}
	\put(-120,75){(\textit{n})}
	\includegraphics[scale = 1] {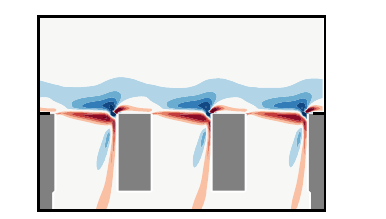}
	\put(-120,75){(\textit{o})} 
	\\
	\includegraphics[scale = 1] {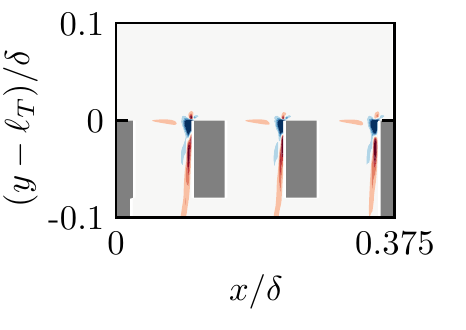}
	\put(-135,100){(\textit{p})}
	\includegraphics[scale = 1] {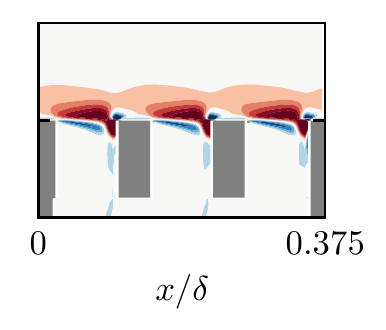}
	\put(-120,100){(\textit{q})}
	\includegraphics[scale = 1] {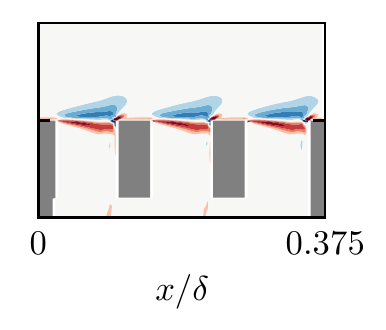}
	\put(-120,100){(\textit{r})} 
	\\
	\includegraphics[scale = 1] {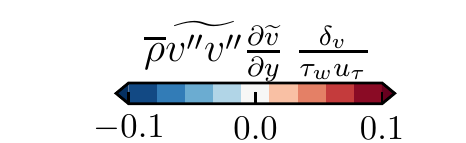}
	\includegraphics[scale = 1] {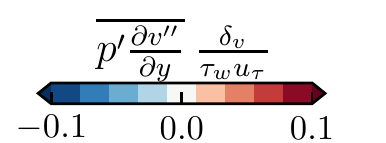}
	\includegraphics[scale = 1] {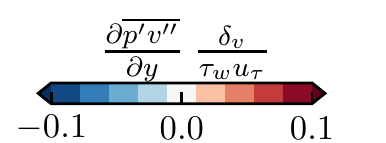} \\
	\caption{Contours of $ \overline{\rho} \widetilde{v''v''} \partial \widetilde{v}/\partial y $ (left),  $\overline{p' \partial v''/\partial y}$ (middle) and $\partial (\overline{p'v''}) /\partial y$ (right), normalised by $\delta_v$, $u_\tau$ and $\tau_w$ for cases $L_1$--$L_6$ (top to bottom).}
	\label{fig:vtke}
\end{figure}

Figure \ref{fig:mean_vtke} shows the budget of the wall-normal turbulent kinetic energy for flow cases $L_5$ and $L_6$ above the wall.
The two dominant terms in the budget appear to be the pressure-strain correlation, $\Pi_{\alpha \alpha}$ and the transport term, $T_k$.
Furthermore, $T_k$ consists primarily of the transport of wall-normal velocity fluctuations via pressure fluctuations.
Energy is redistributed into the wall-normal fluctuations via the pressure-velocity correlation and then also transported with the aid of pressure fluctuations.
A better insight into the wall-normal turbulent kinetic energy budget can be obtained with the help of figure \ref{fig:vtke}, 
which shows a subset of the terms of equation \eqref{eq:vtke2}.
Other than a small region towards the downstream edge of the orifice, 
which corresponds, approximately with the strip of high wall-normal velocity fluctuations seen at the downstream edge in figure \ref{fig:pa},
production of wall-normal velocity fluctuations inside the cavity plays a relatively minor role, 
paling in comparison to the redistribution of turbulent kinetic energy into $v'^2$. 
Pressure fluctuations at the downstream edge redistribute energy into the wall-normal turbulent kinetic energy and this effect appears to be stronger as $d^+$ increases. 
The location of maximum turbulent kinetic energy redistribution corresponds to the location of the peak in wall-normal velocity fluctuations inside the cavity, shown in figure \ref{fig:pa} and \ref{fig:vrms}.

Production plays a minor role and energy is not extracted from the mean flow for the wall-normal velocity fluctuations.
Our findings match those of \citet{yuan_roughness_2014}, who noted in their simulation of sand-grain roughness that 
energy is redistributed into the wall-normal velocity fluctuations that then distort the near wall streamwise structures.

\subsection{Spectral Densities}

\begin{figure}
    \centering
        \includegraphics[scale = 1,trim={0 1cm 0 0},clip] {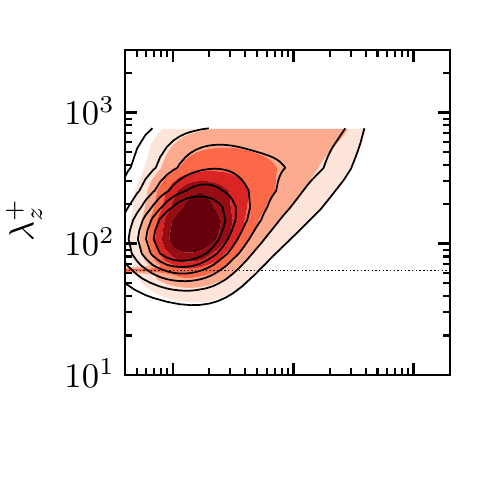} 
        \put(-130,112){(\textit{a})}
        \includegraphics[scale = 1,trim={1cm 1cm 0 0},clip] {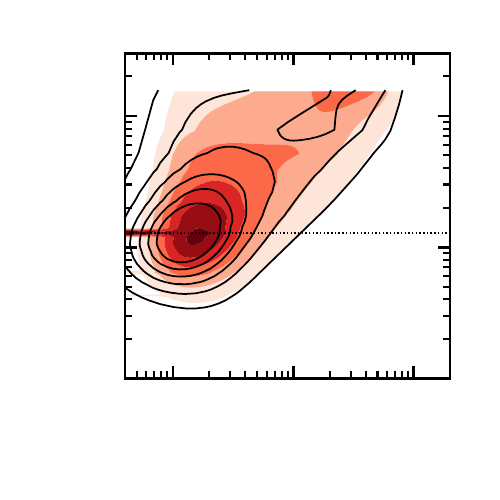} 
        \put(-120,112){(\textit{b})}
        \includegraphics[scale = 1,trim={1cm 1cm 0 0},clip] {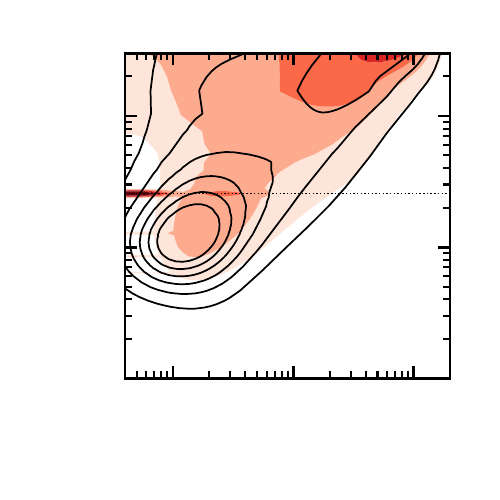} 
        \put(-120,112){(\textit{c})}\\
        \includegraphics[scale = 1] {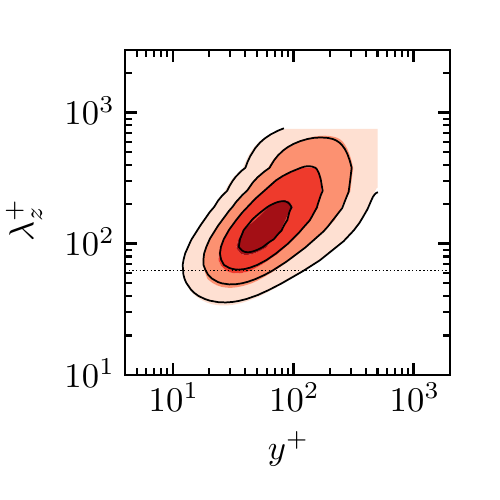} 
        \put(-130,140){(\textit{d})}
        \includegraphics[scale = 1,trim={1cm 0 0 0},clip] {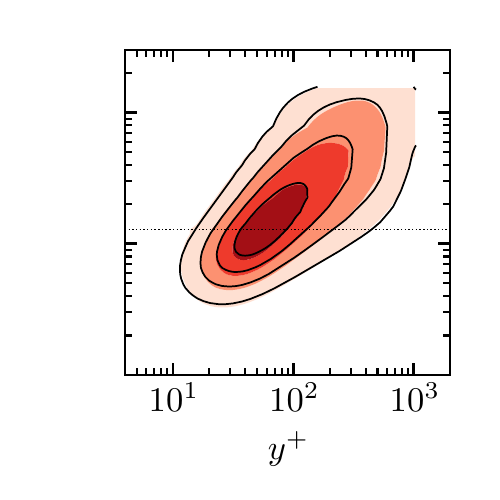} 
        \put(-120,140){(\textit{e})}
        \includegraphics[scale = 1,trim={1cm 0 0 0},clip] {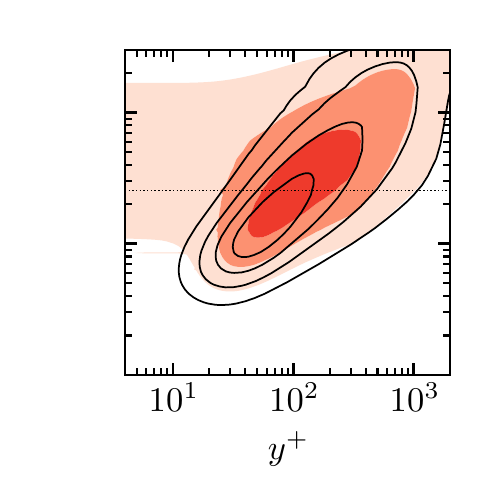} 
        \put(-120,140){(\textit{f})}\\
    \caption{Streamwise velocity (\textit{a})-(\textit{c}) and wall-normal velocity (\textit{d})-(\textit{f}) spectra. Filled contours represent flow case $L_3$ (\textit{a}),(\textit{d}), flow case $L_5$ (\textit{b}),(\textit{e}) and flow case $L_6$ (\textit{c}),(\textit{f}). Contour lines represent smooth wall flow cases at matching $Re_\tau$. The dotted line indicates the spacing of the orifices, normalised by the viscous length scale. Contour levels [1.0,2.0,3.0,4.0,5.0,6.0] are shown for the streamwise velocity spectra and [0.25,0.50,0.75,1.00] are shown for the wall-normal velocity spectra.}
    \label{fig:spectra_z}
\end{figure}

The spectral analysis of velocity fluctuations provides additional insight into the organization of turbulence.
Figures \ref{fig:spectra_z}
show the pre-multiplied spectral densities of the streamwise
and wall-normal velocity components as a function of the wall distance
and the spanwise wavelength $\lambda_z$,
for flow cases $L_3$, $L_5$ and $L_6$.

Spectrograms of the smooth wall cases (isolines) 
show the typical organization that characterizes wall turbulence.
At low Reynolds numbers a near-wall energy peak is evident, at wavelength $\lambda_z^+\approx 100$, corresponding to the near-wall cycle. 
Flow case $S_3$ shows a secondary energy peak in the outer layer for the streamwise velocity fluctuations,
which is associated with the emergence of large-scale energy-containing structures
in the outer layer~\citep{hutchins_07}.

At the lowest Reynolds number, the spectra for the acoustic liners (coloured contour)
match very well the smooth wall results, with minor
differences only visible around the near wall peak.
An obvious difference from the smooth wall is
the presence of a distinct energy
peak in the spanwise velocity spectrogram at a wavelength
corresponding to the spacing of the orifices.
This energy maximum is accompanied
by a decrease of the inner
energy peak, which 
was also observed in the streamwise 
velocity variance in figure~\ref{fig:rs},
and it becomes more prominent 
for increasing Reynolds number.

The same trend is also visible for the wall-normal velocity fluctuations. However, in this case, the near-wall energy peak is visible primarily for the high Reynolds case $L_6$ and
is spread out over a larger band of wavelengths.
We further note that the peak of the near wall cycle is slightly biased towards the length scale corresponding to 
orifices spacing. This wavelength bias was also observed
by~\citet{chu_transport_2021} on porous beds formed by cylindrical elements.

The picture that emerges is
that acoustic liners drain energy 
from the near wall cycle, and tend to re-arrange it 
at length scales typical of the underlying surface pattern,
such as the orifice spacing. This behaviour has also
been reported for other types of surface patterns, 
such as plant canopies~\citep{finnigan_00}.

Away from the wall, the contours of the acoustic liner spectrograms match
fairly well the smooth wall isolines in figure \ref{fig:spectra_z}, suggesting
a similar organization of turbulence in the outer layer, once
the effect of the virtual origin shift is accounted for.
A notable difference is visible for flow case $L_6$,
which is characterized by higher energy levels close to the wall, at large wavelengths.
This may be associated with the footprint
of large-scale structures interacting with the near-wall turbulence.
This behaviour is also typical of flows over smooth walls,
although it emerges at much higher Reynolds numbers~\citep{mathis_09}.
Modification of inner/outer layer interaction
due to surface roughness has been observed in many other studies, 
in different forms.
\citet{efstathiou_mean_2018} observed
an enhancement of the interaction over porous surfaces, 
as compared to the smooth wall, and associated
it to the appearance of spanwise-elongate structures close to the wall. \citet{kim_experimental_2020} also observed stronger
inner/outer layer interaction over porous surfaces, 
and they associated it to the enhanced wall-normal turbulent
mixing caused the relaxation of the impermeability condition.
A similar effect was also reported over rough walls \citep{wu_modelling_2019}. 

It seems that acoustic liners promote inner/outer layer interaction 
at lower Reynolds number, probably because
the flow is approaching the fully rough regime,
thus the viscous effects which would normally mask
this interaction are less prominent and we observe
flow features of high Reynolds number turbulence already at $\Rey_\tau\approx 2000$.
A different interpretation of the same mechanism
can be that the near wall-cycle penetrates deeper
into the porous media, thus effectively reducing the viscous
sublayer perceived by the large-scale eddies, 
which therefore scrapes the surface of the liner at a lower friction Reynolds number than on a smooth wall,
which is also consistent with the interpretation of~\citet{kim_experimental_2020}.

\section{Final Comments}

We performed pore-resolved direct numerical simulations of turbulent flows over perforated plates, which
closely resemble the geometry of acoustic liners in aircraft engines.
This numerical methodology provided us with unprecedented high-fidelity data, 
allowing us to address several aspects of the flow physics towards a
fundamental understanding of turbulent flows over porous surfaces.

Porous surfaces have been studied considerably less than rough surfaces, and the present data
constitute one of the few examples of pore-resolved simulations at high Reynolds numbers.
Acoustic liners induce an increase in drag compared to a smooth wall, 
however, outer layer similarity for the mean velocity is preserved.
We find convincing evidence of a fully rough regime over porous surfaces,
and that the Forchheimer permeability
is the characteristic length scale for acoustic liners because inertia dominates the flow inside the orifices.
This aspect is particularly interesting and novel. 
Most previous studies on porous surfaces
assume that the Darcy law is valid, but this might not be the case for moderate pore sizes.
Darcy-type models have been used extensively to avoid solving the porous surface geometry,
whereas the present findings reveal that this modelling assumption might be inaccurate if the
viscous-scaled pore size is large, and nonlinear corrections for the pressure drop might be needed.
We also believe, that the existence of a fully rough regime should not
be taken for granted for all porous surfaces, as several different geometries
fall into this classification, and not all of them might give rise to the same flow physics.

As for the effect of acoustic liners on turbulence, we observed very high velocity fluctuations inside the orifices.
We find a very strong correlation between the Hama roughness function and the maximum vertical velocity fluctuations for all considered liner goemetries.
Even though this correlation has little relevance from a practical perspective because in general $\widetilde{v''v''}$ is not known,
it shows that flow inside the orifices is well mixed, which explains the success of the Forchheimer permeability.

From an engineering perspective, the existence of a fully rough regime together with outer layer similarity is good news;
together, they form a solid background for wall models and, in principle, we can give up the detailed representation of the surface pattern because the mean velocity profile
presents universal features that can be modelled.
Moreover, these results tell us that it is
possible to use simulations and experiments to estimate the drag variation at higher Reynolds numbers, typical of
practical configurations. 
Our estimate, based on the present DNS,
is that a typical acoustic liner produces about 70\% higher drag per plane area as compared to a hydraulically smooth wall. 
However, this figure does not account for incoming sound waves, whose effects on drag are yet to be understood.\\

{\bf Acknowledgments}
We acknowledge PRACE for awarding us access to Piz Daint, at the Swiss National Supercomputing Centre (CSCS), Switzerland.

\vspace{24pt}

{\bf Declaration of Interests}
The authors report no conflict of interest.

\appendix
\section{Verfication of Numerical Setup}
\begin{figure}
    \centering
        \includegraphics[scale = 1] {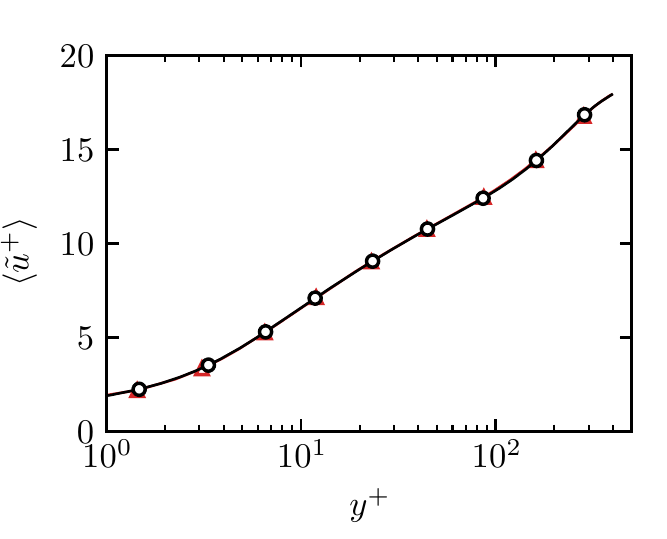} 
        \put(-180,145){(\textit{a})}
        \includegraphics[scale = 1] {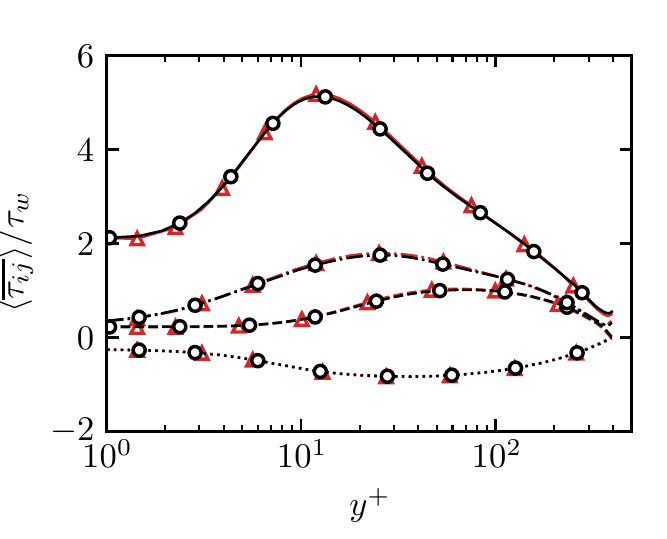} 
        \put(-180,145){(\textit{b})}
        \caption{Comparison of the average streamwise velocity (\textit{a}) and Reynolds stresses (\textit{b}) between STREAmS \citep[triangles]{bernardini_streams_2021} and the DNS of spanwise-aligned bars of \citet[circles]{macdonald_direct_2018} with streamwise spacing $s^+ = 200$ and height $k^+=50$.  In (\textit{b}), different lines represent different components of the Reynolds stress tensor: $\langle \overline{\tau_{11}} \rangle$ (solid), $\langle \overline{\tau_{22}} \rangle$ (dashed), $\langle \overline{\tau_{33}} \rangle$ (dashed-dotted) and $\langle \overline{\tau_{12}} \rangle$ (dotted). }
    \label{fig:macdonald}
\end{figure}

The immersed boundary method has been validated by reproducing the results of~\citet{macdonald_direct_2018}, who performed DNS of open channel flow over spanwise-aligned bars using a body-fitted solver. We reproduced 
this configuration by simulating the flow
over the same roughness geometry, matching the viscous-scaled spacing in the streamwise direction $s^+ = 200$, the roughness height $k^+ = 50$, and the friction Reynolds number $Re_\tau \approx 395$, using grid size $N_x \times N_y \times N_z = 800 \times 250 \times 68$.  
Figure \ref{fig:macdonald} shows a comparison of the mean streamwise velocity and the Reynolds stresses.
Perfect agreement is observed between the DNS of \citet{macdonald_direct_2018} and the present data, 
confirming the accuracy of our immersed boundary method.

\FloatBarrier
\bibliographystyle{jfm}
\bibliography{references}

\end{document}